\documentclass[12pt]{article}
\usepackage{amsmath}
\usepackage{epsfig}
\usepackage{psfrag}
\usepackage{cite}

\def\no{\nonumber}
\def\eps{\varepsilon}
\def\Li{\text{Li}}
\def\S{\text{S}}
\def\as{\alpha_s}
\def\MSbar{\ensuremath{\overline{\text{MS}}}}
\def\calO{\mathcal{O}}
\def\calH{\mathcal{H}}
\def\ub{\bar{u}}

\def\be{\begin{equation}}
\def\ee{\end{equation}}
\def\beq{\begin{eqnarray}}
\def\eeq{\end{eqnarray}}
\def\no{\nonumber}

\def\eps{\epsilon}

\newcommand{\SCETI}{\mbox{SCET${}_{\rm I}$}}

\newcommand{\lqcd}{\Lambda_\mathrm{QCD}}

\newcommand{\spp}{\vphantom{\bigg(}}

\allowdisplaybreaks

\begin{document}


\begin{titlepage}

\begin{flushright}
 TTK-10-42\\
 SI-HEP-2010-10\\
 TTP10-32\\
 SFB/CPP-10-63\\[0.1cm]
 \today
\end{flushright}
\vskip 1.0cm

\begin{center}
\Large{\bf\boldmath
Heavy-to-light currents at NNLO in SCET\\ 
and semi-inclusive 
$\bar B\to X_s\ell^+\ell^-$ decay
\unboldmath}

\normalsize
\vskip 1.5cm

{\sc G.~Bell}$^{a}$,
{\sc M.~Beneke}$^{b}$,
{\sc T.~Huber}$^{c}$
and
{\sc Xin-Qiang~Li}$^{d,e}$\\

\vskip .5cm

{\it $^a$
Albert Einstein Center for Fundamental Physics,\\
Institute for Theoretical Physics,
University of Bern,\\ 
Sidlerstrasse 5, 3012 Bern, Switzerland}\\[0.2cm]
{\it $^b$
Institut f\"ur Theoretische Teilchenphysik 
und Kosmologie,\\
RWTH Aachen University,\\
D--52056 Aachen, Germany}\\[0.2cm]
{\it $^c$
Theoretische Physik 1,
Fachbereich 7, \\
Universit\"at Siegen,
D-57068 Siegen,
Germany}\\[0.2cm]
{\it $^d$
Department of Physics, Henan Normal University,\\
Xinxiang, Henan 453007, P.~R. China}\\[0.2cm]
{\it $^e$
Institute of Theoretical Physics, \\
Chinese Academy of Science,\\
Beijing 100190, P.~R. China}

\vskip 1.8cm

\end{center}

\begin{abstract}
\noindent We perform the two-loop matching calculation for heavy-to-light
currents from QCD onto soft-collinear effective theory
for the complete set of Dirac structures. The newly obtained matching
coefficients enter several phenomenological applications, of which
we discuss heavy-to-light form factor ratios and exclusive radiative 
decays, as well as the semi-inclusive decay 
$\bar B \to X_s \ell^+\ell^-$. For this decay, we observe a significant 
shift of the forward-backward asymmetry zero and find 
$q_0^2 = (3.34 {}^{+0.22}_{-0.25})\,$GeV$^2$
for an invariant mass cut $m_X^{\rm cut}=2.0\,$GeV.

\end{abstract}

\vfill

\end{titlepage}


\section{Introduction}


The flavour-changing quark currents $\bar q \,\Gamma_i\, b$, with 
$\Gamma_i=\{1, \gamma_5, \gamma^{\mu}, \gamma_5 \gamma^{\mu}, 
i\sigma^{\mu\nu}\}$, govern the hadro\-nic dynamics in semi-leptonic 
and radiative $B$ decays. The matrix elements of the currents, usually 
parameterized by several transition form factors, are also important 
inputs to the factorization formulae for non-leptonic 
$B$ decays~\cite{Beneke:1999br}. In the kinematic region where the
hadronic final state has small invariant mass but large energy,  
soft-collinear effective theory~(SCET)~\cite{Bauer:2000yr,Beneke:2002ph} 
is the appropriate theoretical framework, with 
which transparent factorization formulae for the heavy-to-light 
form factors have been derived \cite{Beneke:2003pa} (see
also~\cite{Bauer:2002aj,Lange:2003pk}). Thus, the accurate representation
of the heavy-to-light currents in SCET is of particular interest.

The LO and NLO matching coefficients for heavy-to-light
currents from QCD onto SCET for an arbitrary Dirac matrix has been 
worked out a few years ago~\cite{Bauer:2000yr,Beneke:2000wa,Beneke:2004rc}. 
The coefficients for V-A currents have recently been determined to NNLO
in the context of inclusive semi-leptonic $B$ 
decays~\cite{Bonciani:2008wf,Asatrian:2008uk,Beneke:2008ei,Bell:2008ws}
in the shape-function region.
In this paper we complete the NNLO calculation by computing the 
remaining matching coefficients of the tensor currents. 
The tensor matching coefficients enter several phenomenological 
applications, of which we shall discuss heavy-to-light form factor 
ratios and exclusive radiative decays, as well as the 
semi-inclusive decay $\bar B \to X_s \ell^+\ell^-$.

The paper is organized as follows. In Section~\ref{sec:calculation} 
we first set up notation and then briefly recapitulate the 
techniques applied and the necessary ingredients for the 
two-loop calculation. In Section~\ref{sec:result} the two-loop 
calculation of the QCD form factors and the corresponding matching 
coefficients are presented in detail. Three interesting 
phenomenological applications of our results to heavy-to-light 
form factor ratios, exclusive radiative decays, 
as well as the inclusive decay $\bar B \to X_s \ell^+\ell^-$ are 
discussed in Sections~\ref{sec:application} and \ref{sec:inclusive}. 
We conclude in Section~\ref{sec:conclusion}. The lengthy analytic expressions 
for the coefficient functions can be found in Appendices A and B.


\section{NNLO calculation}
\label{sec:calculation}


\subsection{Set-up of the matching calculation}
\label{sec:setup}

A generic heavy-to-light current $\bar q \,\Gamma_i\, b$ is represented in 
SCET by a set of non-local ``two-body'' and ``three-body'' 
\cite{Beneke:2002ph,Beneke:2003pa,Bauer:2002aj,Lange:2003pk} operators,
\begin{eqnarray}
[\bar q \,\Gamma_i\, b] (0) &=&
\sum_{j} \int ds \; \tilde{C}_i^j(s)
\left[\bar{\xi}W_{hc}\right](sn_+)
\, \Gamma'_j \, h_v(0) \nonumber\\
&& +\, 
\sum_{j} \int ds_1 ds_2 \; \tilde{C}_{i\mu}^{(B1)j}(s_1,s_2)\,
O^{(B1)j\mu}_i(s_1,s_2)
+\ldots, 
\label{eq:matching}
\end{eqnarray}
where $h_v$ is the static heavy quark field defined in HQET, whereas
$\xi$ and $W_{hc}$ are the hard-collinear light quark field and a
hard-collinear Wilson line from SCET, respectively. In this paper 
we are concerned with the calculation of the matching coefficients 
in the first line of (\ref{eq:matching}). The three-body 
operators $O^{(B1)j\mu}_i(s_1,s_2)$ in the second line
are $1/m_b$-suppressed but 
relevant at leading power for exclusive transitions and form factors 
due to the matrix element suppression of the leading term. 
Their one-loop matching 
coefficients are known from \cite{Beneke:2004rc,Hill:2004if} 
and this suffices to work out their contribution to the 
exclusive transitions at ${\cal O}(\alpha_s^2)$. We refer to 
\cite{Beneke:2005gs} for the details of the calculation of 
these spectator-scattering 
terms. In the current work we consider the missing  
${\cal O}(\alpha_s^2)$ matching coefficients of the 
two-body operators $\left[\bar{\xi}W_{hc}\right](sn_+)
\, \Gamma'_j \, h_v(0)$ and adopt the momentum space representation, 
which follows from
\begin{align}
C_{i}^j(n_+p) = \int ds \; e^{is n_+ p} \, \tilde{C}_{i}^j(s).
\end{align}
We decompose the heavy-to-light currents in the basis 
from~\cite{Beneke:2005gs} (summarized in Table~\ref{tab:basis}) 
with two reference vectors $v$ and $n_-$ that fulfill 
$v=(n_-+n_+)/2$, $n^2_\pm=0$ and $n_+n_-=2$. The matching 
calculation involves 12 coefficient functions $C_i^j$, which are 
not independent in a renormalization scheme with anti-commuting 
$\gamma_5$ due to the chiral symmetry of QCD. In the NDR scheme 
adopted in this work, this translates into the constraints 
$C_P=C_S$ and $C_A^i=C_V^i$, while similar relations hold 
between the matching coefficients of the tensor and the 
pseudotensor current. As the latter is reducible in four 
space-time dimensions, we obtain the additional constraints 
$C_T^2=C_T^4=0$ in four dimensions. We nevertheless keep the more 
general basis from Table~\ref{tab:basis}, since we work in 
dimensional regularization and obtain inter\-mediate results that 
are valid in $d=4-2\eps$ dimensions, where $C_T^2$ and $C_T^4$ 
are of $\calO(\eps)$ but non-vanishing.

\begin{table}[b!]
\centerline{
\parbox{15.6cm}{
\renewcommand{\arraystretch}{1.15}
\setlength{\tabcolsep}{2.3mm}
\setlength{\doublerulesep}{0.1mm}
\centerline{
\begin{tabular}{|c||c|c|c|c|c|c|c|c|c|c|c|c|}
\hline
$\Gamma_i$ &
$1$ &
$\gamma_5$ &
\multicolumn{3}{c|}{$\gamma^\mu$} &
\multicolumn{3}{c|}{$\gamma_5 \gamma^\mu$} &
\multicolumn{4}{c|}{$i\sigma^{\mu\nu}$}
\\[0.1em]
\hline
$\Gamma_j'$ &
$1$ &
$\gamma_5$ &
$\gamma^\mu$ & $v^\mu$  & $n_-^\mu$ &
$\gamma_5\gamma^\mu$ & $v^\mu\gamma_5$  & $n_-^\mu\gamma_5$ &
$\gamma^{[\mu}\gamma^{\nu]}$ & $v^{[\mu}\gamma^{\nu]}$ &
$n_-^{[\mu}\gamma^{\nu]}$ & $n_-^{[\mu}v^{\nu]}$
\\[0.1em]
\hline
$C_{i}^j$ &
$C_S$ &
$C_P$ &
$C_V^1$ & $C_V^2$ & $C_V^3$ &
$C_A^1$ & $C_A^2$ & $C_A^3$ &
$C_T^1$ & $C_T^2$ & $C_T^3$ & $C_T^4$
\\[0.1em]
\hline
\end{tabular}}
\caption{\label{tab:basis}\small
\textit{Matching coefficients $C_i^j$ according to the 
decomposition in~(\ref{eq:matching}) 
($a^{[\mu}b^{\nu]}\equiv a^\mu b^\nu - a^\nu b^\mu$).}}}}
\end{table}

It is convenient to perform the matching calculation with on-shell 
quarks and to use dimensional regularization to regularize ultraviolet
(UV) and infrared (IR) singularities. The SCET diagrams are then 
scaleless and vanish and the computation essentially amounts to a 
two-loop calculation in QCD. We, in particular, introduce an analogous
tensor decomposition to (\ref{eq:matching}) and parameterize the QCD 
result in terms of 12 form factors,
\begin{align}
\langle q(p) | \bar q \,\Gamma_i\, b | b(p_b) \rangle
= \sum_{j} \; F_i^j(q^2) \;
\bar{u}(p) \,\Gamma_j'\, u(p_b),
\label{eq:formfactors}
\end{align}
where $p_b=m_b v$ is the momentum of the heavy quark, 
$p = u m_b n_-/2$ the momentum of the light quark and 
$q^2=(p_b-p)^2=(1-u)m_b^2$ denotes the momentum transfer. Due to 
the absence of loop contributions on the effective theory side, 
the SCET matrix elements are given by the tree level matrix elements 
multiplied by a universal renormalization factor $Z_J$ of the 
SCET current $\left[\bar{\xi}W_{hc}\right] \Gamma'_j h_v$. There 
is thus a one-to-one correspondence between the matching coefficients 
$C_i^j$ and the form factors $F_i^j$, 
\begin{align}
C_i^j=Z_J^{-1} F_i^j.
\label{eq:CijFij}
\end{align}
As the form factors are, however, in general IR-divergent, there exists 
no analogous relation on the form factor level to the four-dimensional 
constraints $C_T^2=C_T^4=0$.

The purpose of our analysis consists in the computation of the matching 
coefficients $C_i^j$ (and the respective form factors $F_i^j$) to NNLO 
in QCD. Whereas the NLO corrections have been worked out 
in~\cite{Bauer:2000yr,Beneke:2000wa,Beneke:2004rc}, the coefficients 
$C_V^i$ and $C_A^i$ have recently been determined to NNLO in the context 
of inclusive semi-leptonic $B$ 
decays~\cite{Bonciani:2008wf,Asatrian:2008uk,Beneke:2008ei,Bell:2008ws}. 
In the current work we complete the NNLO calculation by computing the 
remaining matching coefficients $C_S$, $C_P$ and $C_T^i$. 
The four-dimensional constraints mentioned above, will serve as a 
non-trivial check of our calculation.

\subsection{Technical aspects of the calculation}

We organize the calculation along the strategy that we used in our 
previous works on the V-A current~\cite{Beneke:2008ei,Bell:2008ws}. 
The calculation is based on an automated reduction algorithm, which 
uses integration-by-parts techniques~\cite{Tkachov:1981wb} and the 
Laporta algorithm~\cite{Laporta:2001dd} to express the two-loop diagrams 
(shown in Figure~1 of~\cite{Beneke:2008ei}) in terms of a small set of 
scalar master integrals. The required master integrals are already 
known from the computations in~\cite{Bonciani:2008wf,Asatrian:2008uk,Beneke:2008ei,Bell:2007tv,Huber:2009se}.

Our results will be given in terms of the following set of harmonic 
polylogarithms (HPLs)~\cite{Remiddi:1999ew},
\begin{align}
H(0;x) &= \ln(x),  &
H(0,0,1;x) &= \Li_{3}(x),
\no \\
H(1;x) &= -\ln(1-x), &
H(0,1,1;x) &= \S_{1,2}(x),
\no \\
H(-1;x) &= \ln(1+x), &
H(0,0,0,1;x) &= \Li_4(x),
\no \\
H(0,1;x) &= \Li_2(x), &
H(0,0,1,1;x) &= \S_{2,2}(x),
\no \\
H(0,-1;x) &= -\Li_2(-x), &
H(0,1,1,1;x) &= \S_{1,3}(x),
\no \\
H(-1,0,1;x) &\equiv \calH_1(x), &
H(0,-1,0,1;x) &\equiv \calH_2(x),
\end{align}
where we introduced a shorthand notation for the last two HPLs. 
Whereas the first one can be written in a compact form~\cite{Blumlein:1998if},
\begin{align}
\calH_1(x) & =
\ln(1+x) \Li_2(x) + \frac12 \S_{1,2}(x^2)
- \S_{1,2}(x) - \S_{1,2}(-x),
\end{align}
the second one, $\calH_2(x)= \int_0^x dx' \calH_1(x')/x'$, cannot be 
expressed in terms of Nielsen Polylogarithms and has to be evaluated 
numerically.

The charm quark enters the matching calculation at the two-loop level 
through the gluon self energy which contains closed fermion loops. 
Our analytical results from Sections~\ref{sec:Fij} and~\ref{sec:Cij} 
are valid for massless charm quark, but we also show some numerical 
results in Section~\ref{sec:Cij} that include charm mass effects. 
In this case we formally keep $m_c/m_b$ fixed in the heavy-quark 
expansion, so the coefficients depend non-trivially on the quark 
mass ratio (see Section~5 of~\cite{Bell:2008ws}).

The pure two-loop calculation yields bare form factors $F_i^j$ that are 
UV- and IR-divergent. The UV-divergences are subtracted in a standard 
renormalization procedure, which has been described in detail in our 
previous works~\cite{Beneke:2008ei,Bell:2008ws}. We, in particular, 
renormalize the strong coupling constant in the \MSbar-scheme, whereas 
the quark wave-functions and the $b$-quark mass are renormalized in the 
on-shell scheme. The only difference in the current calculation consists 
in the fact that the scalar and the tensor current have non-vanishing 
anomalous dimensions in contrast to the vector current considered 
in~\cite{Beneke:2008ei,Bell:2008ws}. This gives rise to an additional 
multiplicative counterterm $Z_i^{-1}$ ($i=S,T$). We expand the inverse
\begin{align}
Z_i = 1 + \sum_{k=1}^\infty
\left( \frac{\as^{(5)}}{4\pi} \right)^k
Z_i^{(k)}
\end{align}
in terms of the renormalized coupling constant of a theory with five 
active quark flavours. In the \MSbar-scheme the respective NLO coefficients 
are then given by $Z_{S}^{(1)} = 3C_F/\eps$ and $Z_{T}^{(1)} = -C_F/\eps$ 
for the scalar and the tensor current, respectively. At NNLO the 
counter\-terms can be inferred from~\cite{Nanopoulos:1978hh},
\begin{align}
Z_{S}^{(2)} &=
C_F \bigg\{
\bigg[ \frac92 C_F - \frac{11}{2} C_A + 2 n_f T_F \bigg]
\frac{1}{\eps^2}
+ \bigg[ \frac34 C_F + \frac{97}{12} C_A - \frac53 n_f T_F \bigg]
\frac{1}{\eps}
\bigg\},
\no\\
Z_{T}^{(2)} &=
C_F \bigg\{
\bigg[ \frac12 C_F + \frac{11}{6} C_A - \frac23 n_f T_F \bigg]
\frac{1}{\eps^2}
+ \bigg[ \frac{19}{4} C_F - \frac{257}{36} C_A + \frac{13}{9} n_f T_F
\bigg] \frac{1}{\eps}
\bigg\},
\end{align}
where $n_f=5$ denotes the number of active quark flavours.


\section{Results}
\label{sec:result}


\subsection{Renormalized form factors}
\label{sec:Fij}

We first present our results for the renormalized form factors $F_i^j$, 
which are UV-finite but IR-divergent. It will be convenient to decompose 
the form factors according to \begin{align}
F_{i}^j =
\sum_{k=0}^\infty \left( \frac{\as^{(5)}}{4\pi} \right)^k
F_{i}^{j,(k)},\qquad\qquad
F_{i}^{j,(k)} = \sum_l \; F_{i,l}^{j,(k)} \; \eps^l.
\end{align}
In this normalization the form factors become at tree level
\begin{align}
& F_S^{(0)} =
-2F_T^{1,(0)} =
1,
\no\\
& F_T^{2,(0)}  =
F_T^{3,(0)}  =
F_T^{4,(0)} =
0.
\end{align}
Here and below we do not quote our results for the pseudoscalar and the 
(axial) vector current, since the former are equal to those of the scalar 
current in the NDR scheme, while the latter have already been computed 
before and can be found in~\cite{Bonciani:2008wf,Asatrian:2008uk,Beneke:2008ei,Bell:2008ws}.

{\em One-loop form factors.} 
At NLO we compute the form factors up to terms of $\calO(\eps^2)$. Our 
results are given in terms of a set of coefficient functions $g_i(u)$, which 
we list in Appendix~\ref{app:1loop}. The scalar form factor is IR-divergent 
and becomes (with $q^2=\ub m_b^2$, $\ub=1-u$ and $L=\ln\mu^2/m_b^2$),
\begin{align}
F_{S,-2}^{(1)}(u) &= -C_F,
\no\\
F_{S,-1}^{(1)}(u) &= C_F \bigg( g_0(u) - L\bigg),
\no\\
F_{S,0}^{(1)}(u)  &= C_F \bigg( g_1(u) + \Big[ g_0(u) + 3 \Big] L -
\frac12 L^2\bigg),
\no\\
F_{S,1}^{(1)}(u)  &= C_F \bigg( g_2(u) + g_1(u) L
+ \frac12 \Big[ g_0(u)+ 3 \Big] L^2 - \frac16 L^3\bigg),
\no\\
F_{S,2}^{(1)}(u)  &= C_F \bigg( g_3(u) + g_2(u) L + \frac12 g_1(u) L^2 +
\frac16 \Big[ g_0(u) + 3 \Big] L^3 - \frac{1}{24} L^4\bigg).
\end{align}
The first tensor form factor is also IR-divergent and given by
\begin{align}
F_{T,-2}^{1,(1)}(u) &= \frac{C_F}{2},
\no\\
F_{T,-1}^{1,(1)}(u) &= -\frac{C_F}{2} \bigg( g_0(u) - L\bigg),
\no\\
F_{T,0}^{1,(1)}(u)  &= -\frac{C_F}{2} \bigg( g_4(u) + \Big[ g_0(u) - 1
\Big] L - \frac12 L^2\bigg),
\no\\
F_{T,1}^{1,(1)}(u)  &= -\frac{C_F}{2} \bigg( g_5(u) + g_4(u) L  +
\frac12 \Big[ g_0(u) - 1 \Big] L^2 - \frac16 L^3\bigg),
\no\\
F_{T,2}^{1,(1)}(u)  &= -\frac{C_F}{2} \bigg( g_6(u) + g_5(u) L + \frac12
g_4(u) L^2 + \frac16 \Big[ g_0(u) - 1 \Big] L^3 - \frac{1}{24}
L^4\bigg),
\end{align}
whereas the other tensor form factors are IR-finite at NLO and read
\begin{align}
F_{T,0}^{2,(1)}(u)  &= 0,
\no\\
F_{T,1}^{2,(1)}(u)  &= C_F g_7(u),
\no\\
F_{T,2}^{2,(1)}(u)  &= C_F \bigg( g_8(u) + g_7(u) L\bigg),
\\[0.2cm]
F_{T,0}^{3,(1)}(u)  &= C_F g_9(u),
\no\\
F_{T,1}^{3,(1)}(u)  &= C_F \bigg( g_{10}(u) + g_9(u) L\bigg),
\no\\
F_{T,2}^{3,(1)}(u)  &= C_F \bigg( g_{11}(u) + g_{10}(u) L + \frac12 g_9(u)
L^2\bigg),
\\[0.2cm]
F_{T,0}^{4,(1)}(u)  &= 0,
\no\\
F_{T,1}^{4,(1)}(u)  &= C_F g_{12}(u),
\no\\
F_{T,2}^{4,(1)}(u)  &= C_F \bigg( g_{13}(u) + g_{12}(u) L\bigg).
\end{align}

{\em Two-loop form factors.} 
At NNLO the IR-divergent parts of the form factors can be expressed 
in terms of the one-loop coefficient functions $g_i(u)$. The divergent 
terms of the scalar form factor read
\begin{align}
F_{S,-4}^{(2)}(u)  &= \frac12 C_F^2,
\no\\
F_{S,-3}^{(2)}(u)  &= C_F^2 \bigg(L - g_0(u)\bigg) +
  \frac{11}{4} C_A C_F - n_l T_F C_F,
\no\\
F_{S,-2}^{(2)}(u)  &= C_F^2 \bigg[ L^2 - \bigg(2 g_0(u) + 3\bigg) L
+ \frac12 g_0(u)^2 - g_1(u) \bigg]
+ \frac43 L \, T_F C_F
\no\\
& \quad
+ C_A C_F \bigg[ \frac{11}{6} \bigg(L - g_0(u)\bigg)  - \frac{67}{36} +
\frac{\pi^2}{12}  \bigg]
+ n_l T_F C_F \bigg[ \frac59 - \frac23 \bigg(L - g_0(u) \bigg)  \bigg],
\no\\
F_{S,-1}^{(2)}(u)  &=  C_F^2  \bigg[ \frac23 L^3
- \bigg( 2 g_0(u) + \frac92 \bigg) L^2
- \bigg(2g_1(u) - g_0(u)^2 - 3 g_0(u) \bigg) L
\no\\
& \quad
+ g_0(u) g_1(u) - g_2(u) - \frac{3}{8} + \frac{\pi^2}{2} - 6 \zeta_3
\bigg]
\no\\
& \quad
+  C_A C_F \bigg[ \bigg(\frac{\pi^2}{6} - \frac{67}{18} \bigg) \bigg(L -
g_0(u)\bigg) + \frac{461}{216} - \frac{17\pi^2}{24} + \frac{11}{2}
\zeta_3 \bigg]
\no\\
& \quad
+ n_l T_F C_F \bigg[ \frac{10}{9} \bigg(L - g_0(u)\bigg) -\frac{25}{54} +
\frac{\pi^2}{6} \bigg] + T_F C_F \bigg[ 2L^2 - \frac43 g_0(u) L +
\frac{\pi^2}{9} \bigg],
\end{align}
and for the first tensor form factor we get
\begin{align}
F_{T,-4}^{1,(2)}(u)  &= -\frac14 C_F^2,
\no\\
F_{T,-3}^{1,(2)}(u)  &= -\frac12 C_F^2 \bigg(L - g_0(u)\bigg) -
  \frac{11}{8} C_A C_F +\frac12 n_l T_F C_F,
\no\\
F_{T,-2}^{1,(2)}(u)  &= -\frac 12 C_F^2 \bigg[ L^2 - \bigg(2 g_0(u) - 1\bigg) L
+ \frac12 g_0(u)^2 - g_4(u) \bigg]
- \frac23 L \, T_F C_F
\no\\
& \quad
-\frac 12 C_A C_F \bigg[ \frac{11}{6} \bigg(L - g_0(u)\bigg)  - \frac{67}{36} +
\frac{\pi^2}{12}  \bigg]
-\frac 12 n_l T_F C_F \bigg[ \frac59 - \frac23 \bigg(L - g_0(u) \bigg)  \bigg],
\no\\
F_{T,-1}^{1,(2)}(u)  &= -\frac12 C_F^2  \bigg[ \frac23 L^3
- \bigg( 2 g_0(u) - \frac32 \bigg) L^2
- \bigg(2g_4(u) - g_0(u)^2 + g_0(u) \bigg) L
\no\\
& \quad
+ g_0(u) g_4(u) - g_5(u) - \frac38 + \frac{\pi^2}{2} - 6 \zeta_3
\bigg]
\no\\
& \quad
-\frac12  C_A C_F \bigg[ \bigg(\frac{\pi^2}{6} - \frac{67}{18} \bigg) 
\bigg(L -
g_0(u)\bigg) + \frac{461}{216} - \frac{17\pi^2}{24} + \frac{11}{2}
\zeta_3 \bigg]
\no\\
& \quad
-\frac12  n_l T_F C_F \bigg[ \frac{10}{9} \bigg(L - g_0(u)\bigg) 
-\frac{25}{54} +
\frac{\pi^2}{6} \bigg] -\frac12 T_F C_F \bigg[ 2L^2 - \frac43 g_0(u) L +
\frac{\pi^2}{9} \bigg].
\end{align}
The IR-divergent parts of the other tensor form factors are given by
\begin{align}
F_{T,-1}^{2,(2)}(u)  &= -C_F^2 g_7(u),
\end{align}
and
\begin{align}
F_{T,-2}^{3,(2)}(u)  &= -C_F^2 g_9(u),
\no\\
F_{T,-1}^{3,(2)}(u)  &= C_F^2 \bigg( g_0(u) g_9(u) - g_{10}(u) - 2 g_9(u) L
\bigg),
\end{align}
and
\begin{align}
F_{T,-1}^{4,(2)}(u)  &= -C_F^2 g_{12}(u).
\end{align}
The finite parts of the two-loop form factors involve a new set of 
coefficient functions $h_i(u)$, which we specify in Appendix~\ref{app:2loop}. 
We find
\begin{align}
F_{S,0}^{(2)}(u) &= C_F^2 \bigg[
\frac13 L^4
- \bigg( \frac43 g_0(u) + \frac72 \bigg) L^3
- \bigg( 2g_1(u) - g_0(u)^2 - \frac92 g_0(u) -\frac92 \bigg) L^2
\no\\
& \quad \qquad
- \bigg( 2g_2(u) - 2 g_1(u) g_0(u) - 3 g_1(u)
- \frac34 - \pi^2 + 12 \zeta_3 \bigg) L
+ h_1(u) \bigg]
\no\\
& \quad
+ C_A C_F \bigg[
- \frac{11}{18} L^3
+ \bigg( \frac{11}{6} g_0(u) +\frac{16}{9} + \frac{\pi^2}{6} \bigg)L^2
\no\\
& \quad \qquad
+ \bigg( \frac{11}{3} g_1(u) + \Big( \frac{67}{9} - \frac{\pi^2}{3} \Big)
g_0(u) + \frac{2207}{108} - \frac{17\pi^2}{12} + 11 \zeta_3 \bigg)L
+ h_2(u) \bigg]
\no\\
& \quad
+ n_l T_F C_F \bigg[
\frac29 L^3
- \bigg( \frac23 g_0(u) + \frac{8}{9} \bigg)L^2
- \bigg( \frac43 g_1(u) + \frac{20}{9} g_0(u) + \frac{115}{27} -
\frac{\pi^2}{3} \bigg) L
\no\\
& \quad \qquad
- \frac43 g_2(u) - \frac{20}{9} g_1(u) - \bigg( \frac{20}{27} +
\frac{\pi^2}{3} \bigg) g_0(u) - \frac{541}{324}- \frac{13\pi^2}{18} +
\frac{10}{3} \zeta_3 \bigg]
\no\\
& \quad
+ T_F C_F \bigg[\frac{14}{9} L^3
- \bigg( 2 g_0(u) + 2\bigg) L^2
- \bigg( \frac43 g_1(u) +\frac{10}{3} - \frac{2\pi^2}{9} \bigg)L
+ h_3(u) \bigg],
\label{eq:FS:NNLO:finite}
\end{align}
and
\begin{align}
F_{T,0}^{1,(2)}(u) &= -\frac12 C_F^2 \bigg[
\frac13 L^4
- \bigg( \frac43 g_0(u) -\frac76 \bigg) L^3
- \bigg( 2g_4(u) - g_0(u)^2 + \frac32 g_0(u) - \frac12 \bigg) L^2
\no\\
& \quad \qquad
- \bigg( 2g_5(u) - 2 g_4(u) g_0(u) + g_4(u)
- \frac{35}{4} - \pi^2 + 12 \zeta_3 \bigg) L
+ h_4(u) \bigg]
\no\\
& \quad
-\frac12  C_A C_F \bigg[
- \frac{11}{18} L^3
+ \bigg( \frac{11}{6} g_0(u) -\frac{50}{9} + \frac{\pi^2}{6} \bigg)L^2
\no\\
& \quad \qquad
+ \bigg( \frac{11}{3} g_4(u) + \Big( \frac{67}{9} - \frac{\pi^2}{3} \Big)
g_0(u) - \frac{1081}{108} - \frac{17\pi^2}{12} + 11 \zeta_3 \bigg)L
+ h_5(u) \bigg]
\no\\
& \quad
-\frac12 n_l T_F C_F \bigg[
\frac29 L^3
- \bigg( \frac23 g_0(u) - \frac{16}{9} \bigg)L^2
- \bigg( \frac43 g_4(u) + \frac{20}{9} g_0(u) - \frac{53}{27} -
\frac{\pi^2}{3} \bigg) L
\no\\
& \quad \qquad
- \frac43 g_5(u) - \frac{20}{9} g_4(u) - \bigg( \frac{20}{27} +
\frac{\pi^2}{3} \bigg) g_0(u)
+ h_6(u) \bigg]
\no\\
& \quad
-\frac12 T_F C_F \bigg[\frac{14}{9} L^3
- \bigg( 2 g_0(u) - \frac23 \bigg) L^2
- \bigg( \frac43 g_4(u) -\frac{26}{9} - \frac{2\pi^2}{9} \bigg)L
+ h_7(u) \bigg],
\end{align}
and
\begin{align}
F_{T,0}^{2,(2)}(u) &= C_F^2
\bigg( g_0(u) g_7(u) - g_{8}(u) - 2 g_7(u) L \bigg),
\end{align}
and
\begin{align}
F_{T,0}^{3,(2)}(u) &= C_F^2
\bigg[ - 2 g_9(u) L^2
+ \bigg( 2g_0(u) g_9(u) - g_9(u) - 2g_{10}(u) \bigg) L
+ h_8(u)  \bigg]
\no\\
& \quad
+ C_A C_F \bigg[  \frac{11}{3} g_9(u) L + h_9(u)\bigg]
+ T_F C_F \bigg[  - \frac43 g_9(u) L + h_{10}(u)\bigg]
\no\\
& \quad
+ n_l T_F C_F \bigg[ - \frac43 g_9(u) L -\frac43 g_{10}(u) - \frac{8}{9} g_9(u)
 + \frac{4u}{3\ub^2} \ln(u) + \frac{4u}{3\ub} \bigg],
\end{align}
and
\begin{align}
F_{T,0}^{4,(2)}(u) &= C_F^2
\bigg( g_0(u) g_{12}(u) - g_{13}(u) - 2 g_{12}(u) L \bigg).
\label{eq:FT4:NNLO:finite}
\end{align}

\subsection{Matching coefficients}
\label{sec:Cij}

The matching coefficients $C_i^j$ follow from the above expressions for 
the renormalized form factors $F_i^j$ after multiplication with the inverse 
of the renormalization factor of the SCET current $Z_J$, 
cf.~(\ref{eq:CijFij}). To this end one has to keep in mind that the form 
factors have been computed in QCD with five active quark flavours, while 
$Z_J$ is usually given in SCET with four active flavours. We thus have
\begin{align}
Z_J = 1 + \sum_{k=1}^\infty
\left( \frac{\as^{(4)}}{4\pi} \right)^k
Z_J^{(k)},
\end{align}
with NLO coefficient~\cite{Bauer:2000yr},
\begin{align}
Z_J^{(1)} &=
C_F \left\{ -\frac{1}{\eps^2} - \frac{1}{\eps} 
\bigg( \ln \frac{\mu^2}{u^2 m_b^2} + \frac52 \bigg)
\right\}.
\end{align}
The two-loop anomalous dimension can be deduced from~\cite{Becher:2005pd} 
(see also \cite{Asatrian:2008uk})
\begin{align}
Z_J^{(2)} &=
C_F \left\{ \frac{C_F}{2\eps^4}
+ \bigg[ \bigg( \ln \frac{\mu^2}{u^2 m_b^2} + \frac52 \bigg) C_F + 
\frac{11}{4} C_A - n_l T_F
\bigg]\frac{1}{\eps^3} \right.
\no\\
&\quad
+ \bigg[ \frac12 \Big(\ln \frac{\mu^2}{u^2 m_b^2} + \frac52\Big)^2 C_F
+ \bigg( \frac{\pi^2}{12} - \frac{67}{36}  + \frac{11}{6}
\Big( \ln \frac{\mu^2}{u^2 m_b^2} + \frac52 \Big)
\bigg) C_A
\no\\
&\quad
+ \bigg( \frac{5}{9} - \frac{2}{3} \Big(\ln \frac{\mu^2}{u^2 m_b^2} + 
\frac52\Big) \bigg) n_l T_F
\bigg] \frac{1}{\eps^2}
+ \bigg[ \bigg( \frac{\pi^2}{2} - \frac{3}{8} - 6 \zeta_3 \bigg) C_F
\no\\
&\quad
+ \bigg( \frac{461}{216} - \frac{17\pi^2}{24} + \frac{11}{2} \zeta_3 + \Big(
\frac{\pi^2}{6}  - \frac{67}{18} \Big) \Big( \ln \frac{\mu^2}{u^2 m_b^2} + 
\frac52 \Big) \bigg) C_A
\no\\
&\quad
+ \bigg( \frac{\pi^2}{6}- \frac{25}{54} + \frac{10}{9} 
\Big(\ln \frac{\mu^2}{u^2 m_b^2} + \frac52
\Big) \bigg) n_l T_F\bigg] \frac{1}{\eps} \bigg\},
\end{align}
where $n_l=n_f-1=4$ is the number of active quark flavours in the effective 
theory.

We now expand the matching coefficients in terms of the coupling constant of 
the four-flavour theory as
\begin{align}
C_{i}^j =
\sum_{k=0}^\infty \left( \frac{\as^{(4)}}{4\pi} \right)^k
C_{i}^{j,(k)},
\end{align}
and rewrite~(\ref{eq:CijFij}) up to NNLO, which yields
\begin{align}
C_i^{j,(0)} &= F_i^{j,(0)},
\no\\
C_i^{j,(1)} &= F_i^{j,(1)} - Z_J^{(1)} F_i^{j,(0)},
\no\\
C_i^{j,(2)} &= F_i^{j,(2)} + \delta\alpha_s^{(1)} F_i^{j,(1)}
- Z_J^{(1)} \left( F_i^{j,(1)} - Z_J^{(1)} F_i^{j,(0)} \right)
- Z_J^{(2)} F_i^{j,(0)}.
\end{align}
Notice that the last relation implies a term which stems from the conversion
of the five-flavour to the four-flavour coupling constant,
\begin{align}
\as^{(5)} = \as^{(4)} \left[1+\frac{\as^{(4)}}{4\pi} \, \delta\as^{(1)} + {\cal O}(\as^2)\right]
\end{align}
with  (see also~\cite{Beneke:2008ei,Bell:2008ws} for further details)
\begin{align}
\delta\as^{(1)} = T_F \bigg[ \frac43 \ln \frac{\mu^2}{m_b^2}
+ \bigg( \frac23 \ln^2 \frac{\mu^2}{m_b^2} + \frac{\pi^2}{9} \bigg) \eps
+ \bigg( \frac29 \ln^3 \frac{\mu^2}{m_b^2} + \frac{\pi^2}{9} 
\ln \frac{\mu^2}{m_b^2}
-\frac49 \zeta_3 \bigg) \eps^2 + \calO(\eps^3) \bigg].
\end{align}
At LO the matching coefficients then become
\begin{align}
& C_S^{(0)} =
-2C_T^{1,(0)} =
1,
\no\\
& C_T^{2,(0)}  =
C_T^{3,(0)}  =
C_T^{4,(0)} =
0.
\end{align}
At NLO the matching coefficients are given by the finite terms of the 
one-loop form factors,
\begin{align}
C_S^{(1)}(u) &= F_{S,0}^{(1)}(u),
\no\\
C_T^{1,(1)}(u) &= F_{T,0}^{1,(1)}(u),
\no\\
C_T^{3,(1)}(u) &= F_{T,0}^{3,(1)}(u),
\label{eq:Ci:NLO}
\end{align}
and, in particular, $C_T^{2,(1)} = F_{T,0}^{2,(1)} =0$ and 
$C_T^{4,(1)} = F_{T,0}^{4,(1)} = 0$ in accordance with the four-dimensional 
constraints for the tensor coefficients that we mentioned in 
Section~\ref{sec:setup}. Here and in the following we provide 
the expressions for the matching coefficients in the limit 
$\epsilon\to 0$, since the ${\cal O}(\epsilon)$ terms are not relevant 
in two-loop applications.

At NNLO the matching coefficients are no longer given by the finite 
terms of the respective form factors alone. We now find
\begin{align}
C_{S}^{(2)}(u) &= F_{S,0}^{(2)}(u)
\no\\
& \quad
+ T_F C_F \bigg[ \frac{4}{9} \zeta_3 + \frac{\pi^2}{9} g_0(u)
+ \frac{2}{9} \Big( 6 g_1(u) - \pi^2 \Big) L
+  \Big(2g_0(u)+4\Big) L^2 - \frac{14}{9} L^3 \bigg]
\no\\
& \quad
+ C_F^2 \bigg[ g_3(u) - g_0(u) g_2(u) + \Big( 2 g_2(u) - g_0(u) g_1(u)
\Big) L
\no\\
& \hspace{1.5cm}
+ \frac12 \Big( 3 g_1(u) - g_0(u)^2 - 3 g_0(u) \Big) L^2
+ \Big( \frac56  g_0(u) + 2 \Big) L^3 - \frac{5}{24} L^4 \bigg],
\end{align}
and
\begin{align}
C_{T}^{1,(2)}(u) &= F_{T,0}^{1,(2)}(u)
\no\\
& \quad
-\frac12 T_F C_F \bigg[ \frac{4}{9} \zeta_3 + \frac{\pi^2}{9} g_0(u)
+ \frac{2}{9} \Big( 6 g_4(u) - \pi^2 \Big) L
+  \Big(2g_0(u)-\frac43\Big) L^2 - \frac{14}{9} L^3 \bigg]
\no\\
& \quad
-\frac12 C_F^2 \bigg[ g_6(u) - g_0(u) g_5(u) + \Big( 2 g_5(u) - g_0(u) g_4(u)
\Big) L
\no\\
& \hspace{1.5cm}
+ \frac12 \Big( 3 g_4(u) - g_0(u)^2 + g_0(u) \Big) L^2
+ \Big( \frac56  g_0(u) - \frac23 \Big) L^3 - \frac{5}{24} L^4 \bigg],
\end{align}
and
\begin{align}
C_{T}^{3,(2)}(u) &= F_{T,0}^{3,(2)}(u)
+ T_F C_F \bigg[ \frac43 g_9(u) L \bigg]
\no\\
& \quad
+ C_F^2 \bigg[ g_{11}(u) - g_0(u) g_{10}(u)
+\Big( 2g_{10}(u) - g_0(u) g_9(u) \Big) L + \frac32 g_9(u) L^2 \bigg].
\end{align}
The other tensor coefficients are again found to fulfill the four-dimensional 
constraints
\begin{align}
C_{T}^{2,(2)}(u) &= F_{T,0}^{2,(2)}(u)
- C_F^2 \bigg[ g_0(u) g_7(u) - g_8(u) - 2 g_7(u) L \bigg] = 0,
\no\\
C_{T}^{4,(2)}(u) &= F_{T,0}^{4,(2)}(u)
- C_F^2 \bigg[ g_0(u) g_{12}(u) - g_{13}(u) - 2 g_{12}(u) L \bigg] = 0,
\end{align}
which provides a non-trivial cross check of our calculation.

As a further check of our NNLO results we verified that the matching 
coefficients obey the renormalization group equation,
\begin{align}
\frac{d}{d\ln\mu} C_i^j(u;\mu) =
\bigg[ \Gamma_\text{cusp}(\as^{(4)}) \ln \frac{um_b}{\mu}
+ \gamma^\prime(\as^{(4)}) + \gamma_i(\as^{(5)}) \bigg] C_i^j(u;\mu),
\label{eq:RGE}
\end{align}
which consists of a universal piece related to the renormalization 
properties of the SCET current with
\begin{align}
\Gamma_\text{cusp}(\as^{(4)})=
\sum_{k=1}^\infty \left( \frac{\as^{(4)}}{4\pi} \right)^k
\Gamma_\text{cusp}^{(k)},
\hspace{1.5cm}
\gamma^\prime(\as^{(4)}) =
\sum_{k=1}^\infty  \left( \frac{\as^{(4)}}{4\pi} \right)^k
\gamma'^{(k)},
\end{align}
and a second term that contains the anomalous dimension of the QCD 
current with
\begin{align}
\gamma_i(\as^{(5)}) =
\sum_{k=1}^\infty  \left( \frac{\as^{(5)}}{4\pi} \right)^k
\gamma_i^{(k)}.
\end{align}
The one- and two-loop coefficients needed for the check read
$\Gamma_\text{cusp}^{(1)} = 4 C_F$, $\gamma'^{(1)} = -5 C_F$,
\begin{align}
\Gamma_\text{cusp}^{(2)} &=
C_A C_F \bigg[ \frac{268}{9} - \frac{4\pi^2}{3} \bigg]
- \frac{80}{9} n_l T_F C_F,
\no\\
\gamma'^{(2)} &=
C_F^2 \bigg[ 2 \pi^2 - \frac32  - 24 \zeta_3 \bigg]
+ C_A C_F \bigg[ 22 \zeta_3 - \frac{1549}{54} - \frac{7\pi^2}{6} \bigg]
+ n_l T_F C_F \bigg[ \frac{250}{27} + \frac{2\pi^2}{3} \bigg],
\end{align}
and
\begin{align}
\gamma_S^{(1)} &= 6 C_F, &
\gamma_S^{(2)} &=
C_F \bigg[ 3C_F + \frac{97}{3} C_A-\frac{20}{3} (n_l+1) T_F \bigg],
\no\\
\gamma_T^{(1)} &= -2 C_F, &
\gamma_T^{(2)} &=
C_F \bigg[ 19C_F - \frac{257}{9} C_A+\frac{52}{9} (n_l+1) T_F \bigg].
\end{align}
The twofold structure of (\ref{eq:RGE}) can be used to distinguish the 
scale $\mu$, that governs the renormalization group evolution in SCET, from 
a second scale $\nu$, that is related to the non-conservation of the 
scalar/tensor current in QCD. More explicitly the distinction between 
the scales $\mu$ and $\nu$ can be accounted for by writing
\begin{align}
 C_i^j(u;\mu,\nu) = C_i^j(u;\mu) + \delta C_i^j(u;\mu,\nu),
\end{align}
where the first term on the right-hand side 
refers to the above expressions for the matching 
coefficients, $C_i^j(u;\mu)\equiv C_i^j(u)$, while the latter captures 
the dependence on $\ln (\nu/\mu)$, which vanishes when the two scales 
are not distinguished. Expanding the new contribution as
\begin{align}
\delta C_i^j =
\sum_{k=1}^\infty \left( \frac{\as^{(4)}(\mu)}{4\pi} \right)^k
\delta C_{i}^{j,(k)},
\end{align}
we find
\begin{equation}
\delta C_i^{j,(1)}(u;\mu,\nu) = \gamma_i^{(1)}\, C_i^{j,(0)}
\ln \frac{\nu}{\mu} 
\end{equation}
in NLO, and
\begin{eqnarray}
\delta C_i^{j,(2)}(u;\mu,\nu) &=& \left[\frac{{\gamma_i^{(1)}}^2}{2}
-\gamma_i^{(1)}\beta_0^{(5)}\right]\, C_i^{j,(0)}
\ln^2 \frac{\nu}{\mu} \nonumber\\
&& +\, 
\left[ \left(\gamma_i^{(2)}+\frac{4}{3}\, T_F\, 
\gamma_i^{(1)} \,\ln\frac{\mu^2}{m_b^2}\right)\, C_i^{j,(0)}
+\gamma_i^{(1)}\, C_i^{j,(1)}(u;\mu)
\right]\ln \frac{\nu}{\mu}
\end{eqnarray}
in NNLO. (Here $\beta_0^{(5)}=11 C_A/3 - 4/3 \,T_F n_f$ refers to the
QCD beta-function with $n_f=n_l+1$ flavours.)
Our final results for the matching coefficients with the 
two scales $\mu$ and $\nu$ distinct from each other are provided in 
electronic form in~\cite{file}.

The matching coefficients with the SCET and QCD scale distinct from 
each other can be used for additional cross-checks. The scalar 
coefficient is not independent but can be related to the vector 
coefficients by means of the equations of motion, 
yielding~\cite{Hill:2004if}
\be\label{eq:scalarvector}
C_V^1(u;\mu) + \left(1-\frac{u}{2}\right) \, C_V^2(u;\mu) + C_V^3(u;\mu) = 
\frac{\overline{m}_b(\nu)}{m_b} \, C_S(u;\mu,\nu) \, ,
\ee
where $\overline{m}_b(\nu)$ is the \MSbar~renormalized mass in 
five-flavour QCD. Due to the conservation of the vector current
the left-hand side of~(\ref{eq:scalarvector}),
which happens to be just the coefficient $C^{(A0)}_{f_0}$
from~(\ref{eq:linearcombis}) below, is free of $\nu$. 
Hence the QCD scale must also drop out of the right-hand side.
We checked that our results satisfy~(\ref{eq:scalarvector}). 
An equivalent formulation of~(\ref{eq:scalarvector})
was given in~\cite{Bonciani:2008wf} in terms of a Ward-identity.
Also the tensor coefficients at $u=1$, corresponding to $q^2=0$, 
can be checked against existing results in the literature, since they
enter the $b \to s \gamma$ process. From~\cite{Ali:2007sj} 
(see also~\cite{Asatrian:2006ph}) one can infer the combinations
\be
-2 \, F_T^1(u=1) + \frac{1}{2} F_T^2(u=1) + F_T^3(u=1) \label{eq:Ftensorcheck}
\ee
and
\be
-2 \, C_T^1(u=1;\mu,\nu) + C_T^3(u=1;\mu,\nu) \, . \label{eq:Ctensorcheck}
\ee
The latter equation can again be checked for distinct $\mu$ and 
$\nu$, and both~(\ref{eq:Ftensorcheck})
and~(\ref{eq:Ctensorcheck}) agree with the formulas in~\cite{Ali:2007sj}. 
Note that~(\ref{eq:Ctensorcheck}) is just the coefficient
$C^{(A0)}_{T_1}$ from~(\ref{eq:linearcombis}) at $u=1$.

\begin{figure}[t!]
\centerline{\parbox{14cm}{\centerline{
 \psfrag{u}{\scriptsize $u$}
 \psfrag{CS1}{\scriptsize $C_{S/P}$}
 \includegraphics[width=7cm]{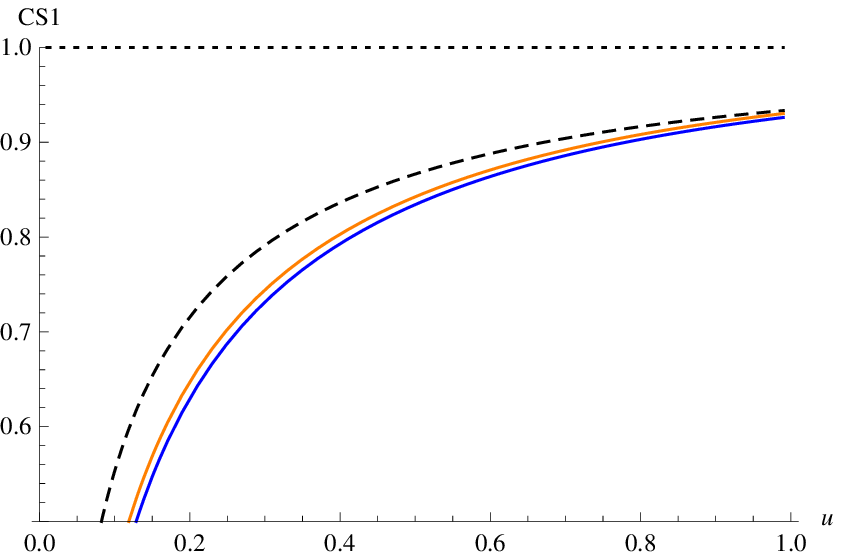}
 \hspace{10mm}
 \psfrag{u}{\scriptsize $u$}
 \psfrag{CV1}{\scriptsize $C_{V/A}^1$}
 \includegraphics[width=7cm]{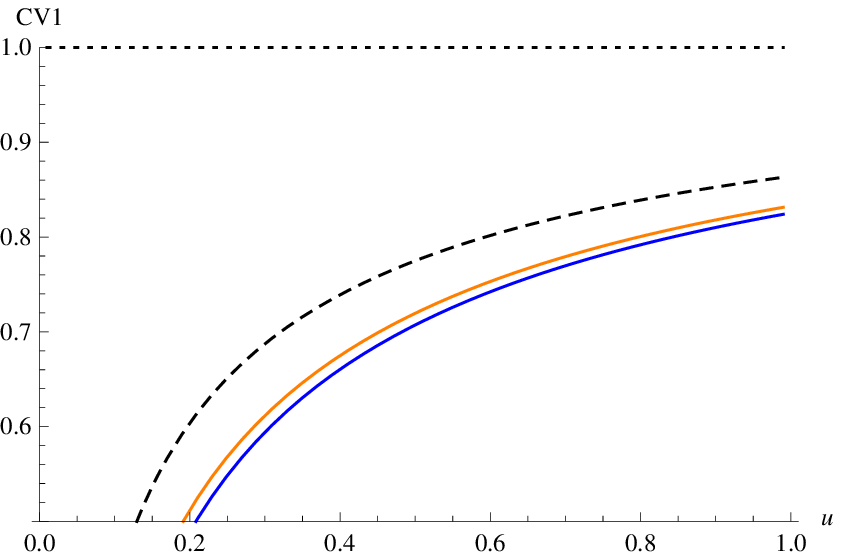}}
\vspace{15mm}
\centerline{
 \psfrag{u}{\scriptsize $u$}
 \psfrag{CV2}{\scriptsize $C_{V/A}^2$}
 \includegraphics[width=7cm]{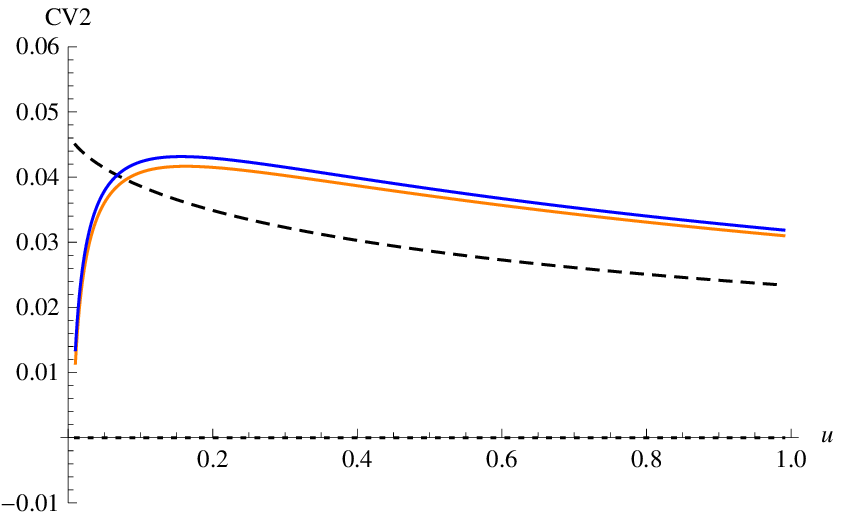}
 \hspace{10mm}
 \psfrag{u}{\scriptsize $u$}
 \psfrag{CV3}{\scriptsize $C_{V/A}^3$}
 \includegraphics[width=7cm]{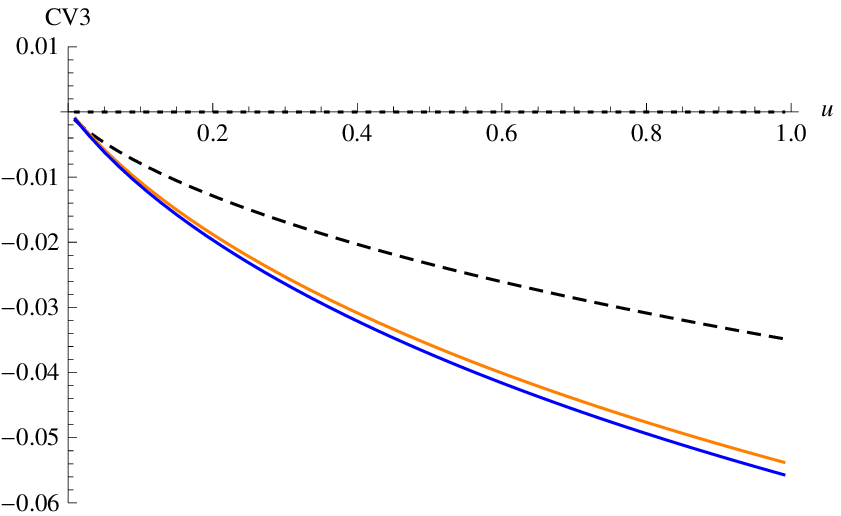}}
\vspace{15mm}
\centerline{
 \psfrag{u}{\scriptsize $u$}
 \psfrag{CT1}{\scriptsize $C_T^1$}
 \includegraphics[width=7cm]{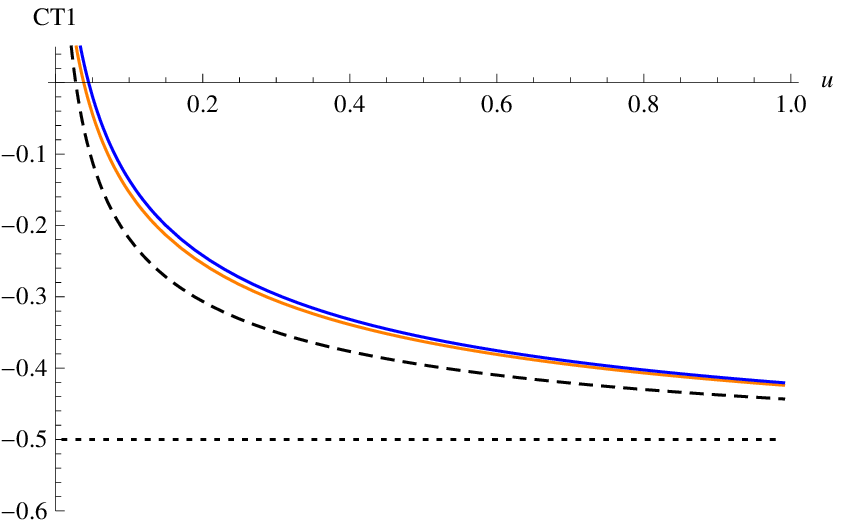}
 \hspace{10mm}
 \psfrag{u}{\scriptsize $u$}
 \psfrag{CT3}{\scriptsize $C_T^3$}
 \includegraphics[width=7cm]{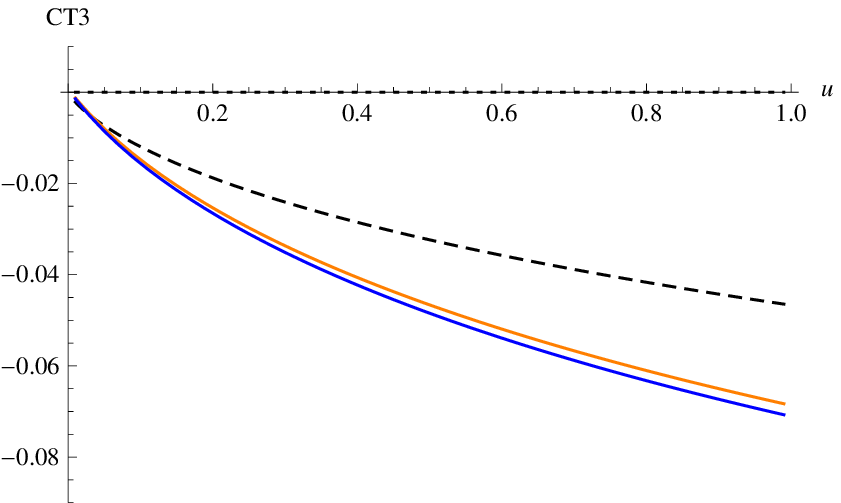}}
\caption{\label{fig:Cij} \small \textit{Matching coefficients 
$C_i^j$ at the scale $\mu=\nu=m_b$ as a function of $u$ 
(the momentum transfer is given by $q^2=(1-u)m_b^2$). 
The dotted horizontal lines show the tree level results, 
the dashed lines the one-loop approximation and the solid 
lines the two-loop approximation with massless charm 
quarks (orange/light grey) and massive charm quarks 
with $m_c/m_b=0.3$ (blue/dark grey).}}}}
\end{figure}

In Figure~\ref{fig:Cij} we evaluate the matching coefficients for 
$\mu=\nu=m_b$ and $\alpha_s^{(4)}(m_b)=0.22$. For completeness we show 
the full set of matching coefficients $C_i^j$ that we
introduced in Table~\ref{tab:basis}. We see that the NNLO corrections 
are in general moderate and add in each case constructively to the NLO 
corrections. In Figure~\ref{fig:Cij} we also
show the effect of a finite charm quark mass,
which is generally rather small, typically modifying the NNLO correction 
by about $10-20\%$.

\section{Exclusive semi-leptonic and radiative $B$ decays}
\label{sec:application}


With the two-loop matching coefficients $C_i^j$ at hand, we explore several 
applications to $B$ meson decays in this and the following section.
For the numerical study we use the following input parameters: the 
$b$-quark pole mass $m_b=4.8\,$GeV; the renormalization scale of the 
QCD scalar and tensor currents $\nu=m_b$; the hard  
scale $\mu=m_b$. The strong coupling constant 
is obtained from $\alpha_s^{(4)}(m_b)=0.215$ by employing three-loop running 
($\Lambda^{(n_f=4)}_{\overline{\rm MS}}=290.9\,$MeV), which gives 
$\alpha_s^{(4)}(1.5\,\mbox{GeV})=0.349$. When we add the 
hard spectator-scattering contribution from \cite{Beneke:2005gs}  as
required for exclusive processes, 
we need further parameters (such as moments of light-cone distribution 
amplitudes), for which we use the same values as 
\cite{Beneke:2005gs} including the hard-collinear scale 
$\mu_{\rm hc}=1.5\,$GeV.

\subsection{Heavy-to-light form factor ratios}
\label{sec:FFratio}

The heavy-to-light form factors in the large-recoil regime, where the 
light meson momentum is parametrically of order of the heavy-quark mass, 
take the following factorization formula~\cite{Beneke:2003pa,Beneke:2000wa}
\begin{eqnarray}\label{eq:factorization}
F_i^{B\to M} (E)\,=\,C_i(E)\,\xi_a(E)+
\int_0^\infty\,\frac{d\omega}{\omega}\,\int_0^1dv\,T_i(E;\ln
\omega,v)\phi_{B+}(\omega)\phi_M(v)\,,
\end{eqnarray}
where $E$ denotes the energy of the light meson $M$, $\xi_a(E)$ is the 
single non-perturbative form factor (one of two when $M$ is a vector meson), 
and $\phi_X$ the light-cone distribution amplitudes of the $B$ meson and 
the light meson. The short-distance coefficients $C_i$ and the 
spectator-scattering kernel $T_i$ can be calculated in perturbation 
theory. The two terms in the above 
equation correspond to the matrix elements of the two terms in the 
operator matching equation (\ref{eq:matching}). In particular, 
the two-loop results from the previous section enter the coefficients 
$C_i(E)$ of the first term. The spectator-scattering kernels $T_i$ 
have been calculated at ${\cal O}(\alpha_s)$ in \cite{Beneke:2000wa}, 
and at ${\cal O}(\alpha_s^2)$ in~\cite{Hill:2004if,Beneke:2005gs}.

In the following we discuss relations between different QCD 
form factors $F_i^{B\to M} (E)$ that can be deduced from the 
factorization formula (\ref{eq:factorization}).
Adopting the same conventions and notations as \cite{Beneke:2005gs}, 
we can express the three independent $B\to P$ form factors as
\begin{eqnarray}
\label{QCDinSCET} f_+(E) &=& C^{(A0)}_{f_+}(E) \,\xi_P(E)+\int d\tau
\,C^{(B1)}_{f_+}(E,\tau) \,\Xi_P(\tau,E)\,,\nonumber\\
 \frac{m_B}{2 E} \,f_0(E) &=& C^{(A0)}_{f_0}(E) \,\xi_P(E)+\int d\tau
\,C^{(B1)}_{f_0}(E,\tau) \,\Xi_P(\tau,E)\,,\nonumber\\
\frac{m_B}{m_B+m_P} \,f_T(E) &=&C^{(A0)}_{f_T}(E) \,\xi_P(E)+\int
d\tau \,C^{(B1)}_{f_T}(E,\tau) \,\Xi_P(\tau,E)\,,
\label{pscet1}
\end{eqnarray}
and the seven independent $B\to V$ form factors as
\begin{eqnarray}
\frac{m_B}{m_B+m_V} \,V(E) &=& C^{(A0)}_{V}(E) \,\xi_\perp(E)+\int
d\tau
\,C^{(B1)}_{V}(E,\tau) \,\Xi_\perp(\tau,E)\,,\nonumber\\
\frac{m_V}{E}\,A_0(E) &=&C^{(A0)}_{f_0}(E) \,\xi_\parallel(E)+\int
d\tau
\,C^{(B1)}_{f_0}(E,\tau) \,\Xi_\parallel(\tau,E)\,,\nonumber\\
\frac{m_B+m_V}{2 E} \,A_1(E) &=&C^{(A0)}_{V}(E) \,\xi_\perp(E)+\int
d\tau
\,C^{(B1)}_{V}(E,\tau) \,\Xi_\perp(\tau,E)\,,\nonumber\\
\frac{m_B+m_V}{2 E} \,A_1(E) &-&\frac{m_B-m_V}{m_B}
\,A_2(E)\nonumber\\
&=&C^{(A0)}_{f_+}(E) \,\xi_\parallel(E)+\int d\tau
\,C^{(B1)}_{f_+}(E,\tau) \,\Xi_\parallel(\tau,E)\,,\nonumber\\
T_1(E) &=&C^{(A0)}_{T_1}(E) \,\xi_\perp(E)+\int d\tau
\,C^{(B1)}_{T_1}(E,\tau) \,\Xi_\perp(\tau,E)\,,\nonumber \\
\frac{m_B}{2 E} \,T_2(E) &=&C^{(A0)}_{T_1}(E) \,\xi_\perp(E)+\int
d\tau
\,C^{(B1)}_{T_1}(E,\tau) \,\Xi_\perp(\tau,E)\,,\nonumber\\
\frac{m_B}{2 E} \,T_2(E) - T_3(E)
&=&C^{(A0)}_{f_T}(E) \,\xi_\parallel(E)+\int d\tau
\,C^{(B1)}_{f_T}(E,\tau) \,\Xi_\parallel(\tau,E).
\label{vscet1}
\end{eqnarray}
Here $m_B$ represents the $B$ meson mass, $m_{P}$ and $m_V$ refer to the 
pseudoscalar and vector light meson masses, respectively. 
The coefficient functions $C^{(A0)}_F$ and $C^{(B1)}_F$ are defined 
as linear combinations of the matching coefficients of two- (``A0-type'') 
and three-body (``B-type'')  
SCET operators, while $\Xi_a(\tau,E)$ denotes the matrix elements 
of the three-body operators $O^{(B1)j\mu}_i(s_1,s_2)$, 
see (\ref{eq:matching}).
In terms of the coefficients $C_i^j$ introduced in previous sections, 
the five independent A0-coefficients are given by 
\begin{eqnarray}\label{eq:linearcombis}
C^{(A0)}_{f_+}&=&C_V^{1}(u;\mu)+\frac{u}{2}C_V^{2}(u;\mu)+C_V^{3}(u;\mu)\,,\nonumber\\
C^{(A0)}_{f_0}&=&C_V^{1}(u;\mu)+\left(1-\frac{u}{2}\right)C_V^{2}(u;\mu)+C_V^{3}(u;\mu)\,,
\nonumber\\
C^{(A0)}_{f_T}&=&-2 C_T^{1}(u;\mu,\nu)+C_T^{2}(u;\mu,\nu)-C_T^{4}(u;\mu,\nu)\,,\nonumber\\
C^{(A0)}_{V}&=&C^{1}_{V}(u;\mu)\,,\nonumber\\
C^{(A0)}_{T_1}&=&-2 C_T^{1}(u;\mu,\nu)+\left(1-\frac{u}{2}\right)C_T^{2}(u;\mu,\nu)+C_T^{3}(u;\mu,\nu)\,.
\end{eqnarray}
Recall that in $D=4$ dimensions one has $C_T^{2}=C_T^{4}=0$. The variable $E$ used in (\ref{pscet1}) and (\ref{vscet1}) is related to 
$u$ through $u=2E/m_B$. The five independent B-coefficients are given in 
Appendix~A2 of~\cite{Beneke:2005gs}.

From (\ref{pscet1}) and (\ref{vscet1}), we have the following two identities
\be
\frac{m_B}{m_B+m_V} \,V(E) = \frac{m_B+m_V}{2 E} \,A_1(E),
\qquad
T_1(E) = \frac{m_B}{2 E} \,T_2(E)
\label{exactrel}
\ee
up to power corrections~\cite{Burdman:2000ku}. In the physical form factor 
scheme~\cite{Beneke:2000wa,Beneke:2005gs}, where the \SCETI{} form factors 
$\xi_a(E)$ are defined in terms of three QCD form factors,
\be
\xi_P^{\rm FF} \equiv f_+,
\qquad
\xi_\perp^{\rm FF} \equiv \frac{m_B}{m_B+m_V} \,V,
\qquad
\xi_{\parallel}^{\rm FF} \equiv
\frac{m_B+m_V}{2 E} \,A_1 -\frac{m_B-m_V}{m_B}
\,A_2,
\label{ffscheme}
\ee
the five remaining form factors read
\beq
\frac{m_B}{2 E} \,f_0 &=&
R_0 \, \xi_P^{\rm FF} +
\left(C^{(B1)}_{f_0}-C^{(B1)}_{f_+} \,R_0\right)
\star \,\Xi_P,
\nonumber \\[0.2cm]
\frac{m_B}{m_B+m_P} \,f_T &=&
R_T\, \xi_P^{\rm FF} +
\left(C^{(B1)}_{f_T}-C^{(B1)}_{f_+}\, R_T\right)
\star \,\Xi_P,
\nonumber\\[0.2cm]
T_1 &=&
R_\perp \, \xi_\perp^{\rm FF} +
\left(C^{(B1)}_{T_1}-C^{(B1)}_{V} \,R_\perp\right)
\star \,\Xi_\perp,
\nonumber \\[0.2cm]
\frac{m_V}{E}\,A_0  &=&
R_0 \, \xi_\parallel^{\rm FF} +
\left(C^{(B1)}_{f_0}-C^{(B1)}_{f_+} \,R_0\right)
\star \,\Xi_\parallel,
\nonumber \\[0.2cm]
\frac{m_B}{2 E} \,T_2 - T_3 &=&
R_T\, \xi_\parallel^{\rm FF} +
\left(C^{(B1)}_{f_T}-C^{(B1)}_{f_+}\, R_T\right)
\star \,\Xi_\parallel.
\label{physscet1}
\eeq
In this scheme there are only three non-trivial ratios $R$ and three 
non-trivial combinations of B-coefficients, defined, respectively, as
\beq
R_0(u)&\equiv& \frac{C^{(A0)}_{f_0}}{C^{(A0)}_{f_+}}
=1+\frac{\alpha_s^{(4)}}{4\pi}\,C_F\,\left[2+g_9(u)\right] 
\left[1+\frac{\alpha_s^{(4)}}{4\pi}\;\beta_0^{(4)} L_\mu \right] \no \\
&& 
\hspace*{-1cm}
+\bigg(\frac{\alpha_s^{(4)}}{4\pi}\bigg)^{\! 2} \, \bigg\{ C_F^2 \, j_1(u)
+ C_F \, C_A \, j_2(u) + C_F \, n_l \, T_F \, j_3(u)+ 
C_F \, T_F \, j_4(u)\bigg\} + {\cal O}(\alpha_s^3) \,\,, \no \\[0.2cm]
R_T(u)&\equiv& \frac{C^{(A0)}_{f_T}}{C^{(A0)}_{f_+}}
=1+\frac{\alpha_s^{(4)}}{4\pi}\,C_F\,\left[-L_{\nu}-g_9(u)\right] 
\left[1+\frac{\alpha_s^{(4)}}{4\pi}\;\beta_0^{(4)} L_\mu \right] \no \\
&&\hspace*{-1cm}
+\bigg(\frac{\alpha_s^{(4)}}{4\pi}\bigg)^{\! 2} \, \left\{ C_F^2 
\!\left[\frac{L_\nu^2}{2} + \left(\!\frac{19}{2}+g_9(u)\!\right) L_\nu+
j_5(u)\right] \!
 + C_F \, C_A \!\left[\frac{11L_\nu^2}{6}-
\frac{257L_\nu}{18}+j_6(u)\right] \right. \no \\
&&
\hspace*{-1cm}
\left. + C_F \, n_l \, T_F \! 
\left[-\frac{2L_\nu^2}{3}+\frac{26L_\nu}{9}+j_7(u)\right] \!
+ C_F \, T_F \! \left[-\frac{2L_\nu^2}{3}+\frac{26L_\nu}{9} +
j_8(u)\right]\right\} + {\cal O}(\alpha_s^3) \,\,, \no \\[0.2cm]
R_\perp(u)&\equiv& \frac{C^{(A0)}_{T_1}}{C^{(A0)}_{V}}
=1+\frac{\alpha_s^{(4)}}{4\pi}\,C_F\,\left[-L_{\nu}+\frac{1}{2}\,g_9(u)\right] 
\left[1+\frac{\alpha_s^{(4)}}{4\pi}\;\beta_0^{(4)} L_\mu \right] \no \\
&&
\hspace*{-1cm}
+\bigg(\frac{\alpha_s^{(4)}}{4\pi}\bigg)^{\! 2} \, \left\{ C_F^2 \! 
\left[\frac{L_\nu^2}{2} + \left(\frac{19}{2}-\frac{1}{2}\, g_9(u)\right) L_\nu
-\frac{1}{2}\,j_5(u)+j_9(u)\right]\right.\no \\
&& 
\hspace*{-1cm}
+ C_F \, C_A \! \left[\frac{11L_\nu^2}{6}-\frac{257L_\nu}{18}-
\frac{1}{2}\,j_6(u)+j_{10}(u)\right] \!
+ C_F \, n_l \, T_F \! \left[-\frac{2L_\nu^2}{3}+\frac{26L_\nu}{9} -
\frac{1}{2}\,j_7(u)\right. \no \\
&& 
\hspace*{-1cm}
\left.\left. +\frac{2\pi^2}{3}+\frac{205}{36}\right] \!
+ C_F \, T_F \! \left[-\frac{2L_\nu^2}{3}+\frac{26L_\nu}{9}-\frac{1}{2}\, 
j_8(u)-\frac{4\pi^2}{3}+\frac{421}{36}\right]\right\} + 
{\cal O}(\alpha_s^3) \,\,,
\label{rfactors}
\eeq
and
\begin{eqnarray}
C_{0+}^{(B1)}(\tau,E)&=&C_{f_0}^{(B1)}(\tau,E)-
C_{f_+}^{(B1)}(\tau,E)\,R_0(E),
\nonumber \\
C_{T+}^{(B1)}(\tau,E)&=&C_{f_T}^{(B1)}(\tau,E)-
C_{f_+}^{(B1)}(\tau,E)\,R_T(E),
\nonumber \\
C_{T_1V}^{(B1)}(\tau,E)&=&C_{T_1}^{(B1)}(\tau,E)-
C_{V}^{(B1)}(\tau,E)\,R_\perp(E).
\label{cb1phys}
\end{eqnarray}
We denote $L_{\mu}=\ln(\mu^2/m_b^2)$, $L_{\nu}=\ln(\nu^2/m_b^2)$, 
and $\beta_0^{(4)} = 11/3 \, C_A - 4/3 \, T_F \, n_l$. 
The functions $j_i(u)$ can be found in
Appendix~\ref{app:2loop}. One recognizes the relatively simple
structure of the ratios $R_X$ in the physical form factor scheme. 
Compared to the matching coefficients,
where we encounter up to the fourth power of logarithms, the ratios 
$R_X$ have logarithmic dependences that are at most quadratic, since 
the universal Sudakov logarithms cancel in the ratios.

\begin{figure}[t!]
\begin{center}
\epsfig{file=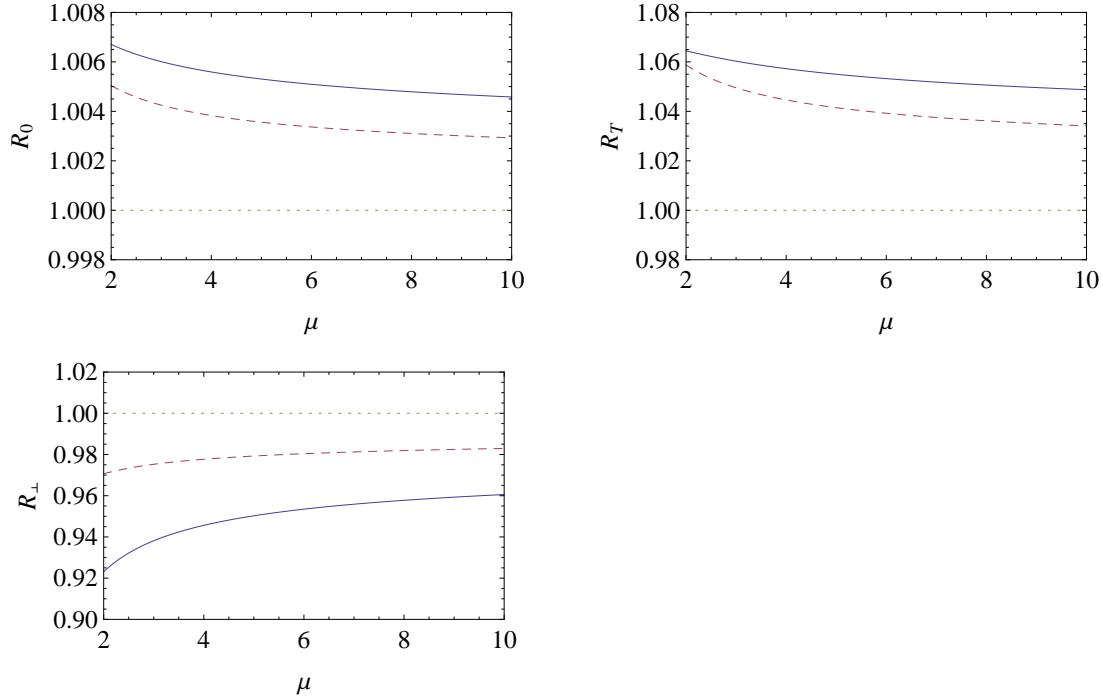,width=15cm}
\caption{\label{fig:scale_dep} 
\small \textit{Dependence of the ratios 
$R_X$~($X=0,T,\perp$) defined in (\ref{rfactors}) 
on the scale $\mu$, with $u=0.85$~(corresponding 
to the light-meson energy $E=u m_B/2 =2.24\,$GeV or momentum 
transfer $q^2=4.18\,$GeV$^2$) and $\nu=m_b$~(the 
renormalization scale of the QCD tensor current). All of 
them equal 1 in the absence of radiative and power 
corrections~(dotted line). The solid and dashed lines 
denote the NNLO and NLO results, respectively.}}
\end{center}
\end{figure}

As expected in any perturbative QCD calculation, the higher-order 
correction is necessary to eliminate scale ambiguities. While the
A0-coefficients $C_X^{(A0)}$ depend on the hard scale $\mu$~(which is 
cancelled by the corresponding dependence of the \SCETI{} form
factors $\xi_a(E)$), the $\mu$ dependence of the ratios 
$R_X$~($X=0,T,\perp$) arises only from the scale-dependence of 
$\alpha_s(\mu)$ and should be reduced after including the higher-order
correction. In Figure~\ref{fig:scale_dep}, we show the dependence of the 
three ratios $R_X$ on the scale $\mu$ at $u=0.85$~(corresponding to the 
light-meson energy $E=u m_B/2 =2.24\,$GeV or momentum transfer 
$q^2=4.18\,$GeV$^2$) and fixed renormalization scale $\nu=m_b$ 
of the QCD tensor current. In the absence of radiative and power
corrections, all these coefficients
equal 1~(dotted lines). We observe that the scale dependence is 
reduced at the two-loop order for the ratios $R_{0,T}$, but not 
for $R_{\perp}$, which receives a large two-loop 
correction.

\begin{figure}[p]
\begin{center}
\epsfig{file=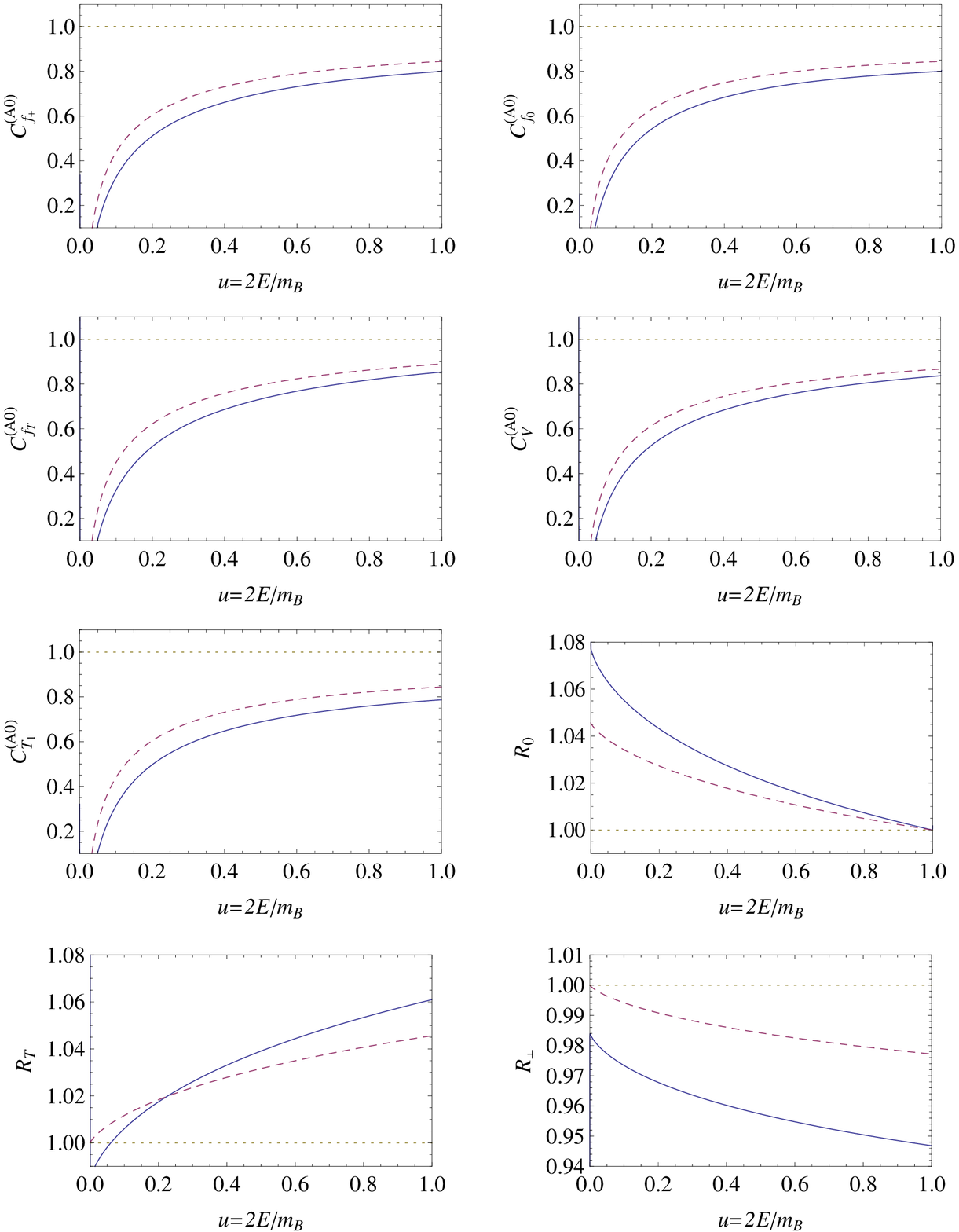,width=15cm}
\caption{\label{fig:qsquare_dep} \small \textit{The A0-type 
coefficients $C_X^{(A0)}$ and the ratios $R_X$~($X=0,T,\perp$) defined in (\ref{rfactors}) as a function of $u$~(related to light-meson energy $E=u\,m_B/2$ or momentum transfer $q^2=(1-u)\,m_B^2$), with the scales fixed at $\nu=\mu=m_b$. The legend is the same as in Figure~\ref{fig:scale_dep}.}}
\end{center}
\end{figure}

Since the A0-type coefficients $C_X^{(A0)}$ and hence the ratios 
$R_X$ also depend on the momentum transfer $q^2$, we show in 
Figure~\ref{fig:qsquare_dep} these coefficients as a function 
of $u$~(related to light-meson energy $E=u\,m_B/2$ or momentum 
transfer $q^2=(1-u)\,m_B^2$), with the scales fixed at $\nu=\mu=m_b$. 
As illustrated in Figure~\ref{fig:qsquare_dep}, the NNLO correction to 
all the five coefficients $C_X^{(A0)}$ is quite similar and adds in each 
case constructively to the NLO result; among the three ratios $R_X$, the 
two-loop correction to $R_{\perp}$, i.e. to the ratio of the tensor 
and vector form factor, $T_1/V$, is most significant.

\begin{figure}[t!]
\begin{center}
\epsfig{file=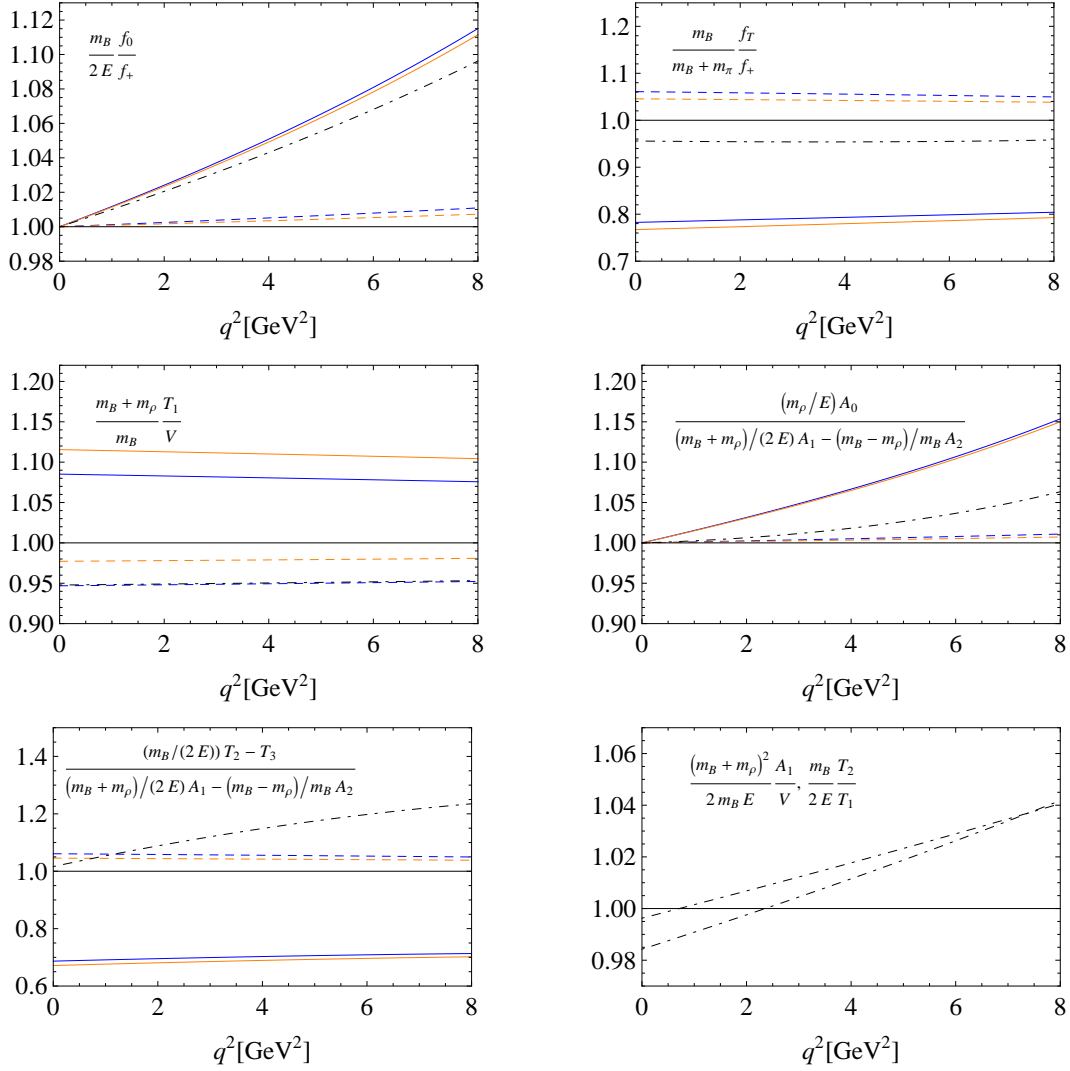,width=15cm}
\caption{\label{fig:ff_ratio} \small \textit{Corrections to the $B\to\pi$ and $B\to\rho$ form factor ratios as a function of momentum transfer $q^2$. All the ratios equal 1 in the absence of radiative corrections. Solid curves: full results with $R_X$ evaluated at NNLO~(blue/dark grey) and NLO~(orange/light grey), including the spectator-scattering term; Dashed: results without the spectator-scattering contribution; Dash-dotted: results from QCD sum rule calculation. The lower right panel shows the two form factor ratios that equal 1 at leading power. For comparison, the QCD sum rule results for these two ratios are also shown~(upper line refers to $A_1/V$, lower line to $T_2/T_1$).}}
\end{center}
\end{figure}

To further investigate these two-loop corrections to the form factor 
ratios, following \cite{Beneke:2005gs} we also take the $B\to\pi$ 
and $B\to\rho$ transitions as examples. Seven ratios among the total 
of ten pion and $\rho$ meson form factors can be obtained from the 
two identities (\ref{exactrel}), which do not receive any perturbative 
corrections, and the five relations that follow from (\ref{physscet1}) 
by dividing through the appropriate $\xi_a^{\rm FF}$. The $q^2$ 
dependence of these form factor ratios are shown in 
Figure~\ref{fig:ff_ratio}. As in \cite{Beneke:2005gs} the $q^2$-dependence 
of the $\xi_a^{\rm FF}$ in the normalization of the spectator-scattering 
correction is taken from the QCD sum rule calculation.
The ratios are normalized such that in absence of any radiative and 
power corrections they equal 1 for all $q^2$. Our final results, 
including both $R_X$ and the spectator-scattering term to order 
$\alpha_s^2$, are shown as solid dark grey (blue in colour) curves, while the 
results with $R_X$ evaluated only at NLO as solid light grey (orange 
in colour) 
ones. One can see that the radiative correction always enhances the 
symmetry-breaking effect, and the NNLO term is generally quite moderate; 
the most significant effect from the two-loop correction is on the ratio 
$T_1/V$~(through the ratio $R_{\perp}$). To see the relative size of 
the two terms in the factorization formula (\ref{eq:factorization}), 
we also show the result without the spectator-scattering term (dashed 
curves with blue/dark grey and orange/light grey denoting the NNLO and 
NLO results, respectively). Comparing the solid with the dashed curves, 
one can see that the radiative correction from the A0-coefficients 
$C_{X}^{(A0)}$ is always smaller than the spectator-scattering contribution.

To compare our results with the QCD sum rule calculations~\cite{Ball:2004ye}, 
the sum rule predictions for these form factor ratios are shown as 
dash-dotted curves in Figure~\ref{fig:ff_ratio}. One notices that, 
while the sum rule calculation generally satisfies the symmetry relations 
better than predicted on the basis of the heavy-quark limit corrected by 
radiative and spectator-scattering effects, see for instance the lower 
right panel of Figure~\ref{fig:ff_ratio}, there are also significant 
differences concerning the sign of the correction, which might be due 
to $1/m_b$ power corrections or ununderstood systematics of the sum 
rule calculations; further detailed discussions could be found 
in \cite{Beneke:2000wa,Beneke:2005gs,DeFazio:2005dx}. The new two-loop 
correction does not affect the conclusions on this point.

\subsection{Exclusive radiative $B$ decays}
\label{sec:exclusive}

As factorization calculations of exclusive radiative and hadronic $B$ 
decays involving only light mesons make use of 
the heavy-to-light form factors at maximal recoil, it is of interest to 
investigate the short-distance corrections at $u=1$, i.e. $E=m_B/2$ or 
$q^2=0$. In this subsection we shall consider the following two 
ratios~\cite{Beneke:2005gs}
\begin{eqnarray}
{\cal R}_1(E) &\equiv& \frac{m_B}{m_B+m_P}\frac{f_T(E)}{f_+(E)}
= R_T(E) +\int_0^1 d\tau \,C_{T+}^{(B1)}(\tau,E)\,
\frac{\Xi_P(\tau,E)}{f_+(E)},
\nonumber\\
{\cal R}_2(E) &\equiv& \frac{m_B+m_V}{m_B}\frac{T_1(E)}{V(E)}=
R_\perp(E)+ \frac{m_B+m_V}{m_B}\int_0^1 d\tau\,
C_{T_1V}^{(B1)}(\tau,E)\,\frac{\Xi_\perp(\tau,E)}{V(E)},\qquad
\label{ffratios}
\end{eqnarray}
defined in the physical form factor scheme.

At $u=1$ and assuming the asymptotic form for the light-meson distribution 
amplitude $\phi_M(v)=6 v\bar v$, the analytic expressions for
these two ratios simplify considerably, even at NNLO. As the 
spectator-scattering contribution is already given by Eq.~(124) in
\cite{Beneke:2005gs}, here we give only the expressions for the ratios 
$R_{T,\perp}$ at $u=1$~(as a consequence of the equations of
motion, we have $R_0(u=1) \equiv 1$),
\begin{eqnarray}
R_T(u=1) &=& 1 + \frac{\alpha_s^{(4)}}{4\pi}\left[\frac{8}{3} - 
\frac{4}{3}L_{\nu}\right]
+\bigg(\frac{\alpha_s^{(4)}}{4\pi}\bigg)^{\! 2} \,
\left[-\frac{100}{9}L_{\mu}L_{\nu} + \frac{200}{9}L_{\mu} + 6L_{\nu}^2 - 
\frac{922}{27}L_{\nu}\, \right. \nonumber\\
&& \left. - \frac{16}{3}\zeta(3) + \frac{10}{3}\pi^4 - \frac{952}{27}\pi^2 + 
\frac{8047}{162} + \frac{128}{27}\pi^2 \ln{2}\,\right]\,,
\nonumber\\[0.2cm]
R_{\perp}(u=1) &=& 1 + \frac{\alpha_s^{(4)}}{4\pi}\left[- \frac{4}{3} - 
\frac{4}{3}L_{\nu}\right]
+\bigg(\frac{\alpha_s^{(4)}}{4\pi}\bigg)^{\! 2} \,
\left[-\frac{100}{9}L_{\mu}L_{\nu} - \frac{100}{9}L_{\mu} + 6L_{\nu}^2 - 
\frac{778}{27}L_{\nu}\, \right. \nonumber\\
&& \left. + 4\zeta(3) - \frac{5}{3}\pi^4 + \frac{428}{27}\pi^2 - 
\frac{13013}{162} - \frac{88}{27}\pi^2 \ln{2}\,\right]\,,
\end{eqnarray}
with $L_{\mu}=\ln(\mu^2/m_b^2)$, $L_{\nu}=\ln(\nu^2/m_b^2)$, and 
$n_l=4$ has been used. Using the three-loop running coupling and 
specifying to the pion (${\cal R}_1$) and $\rho$ meson (${\cal R}_2$), 
numerically we obtain (setting $\nu=\mu=m_b$)
\begin{eqnarray}
{\cal R}_1(E_{\rm max})
&=& 1 + \Big[0.046\,(\mbox{NLO}) + 0.015\,(\mbox{NNLO})\Big]\,(R_T)\, 
\nonumber\\
&& - 0.160\,\Big\{1+0.524\,(\mbox{NLO spec.})
- 0.002 \,(\delta_{\rm log}^\parallel)\Big\}
\nonumber\\[0.2cm]
&=& 0.817,
\nonumber\\[0.4cm]
{\cal R}_2(E_{\rm max})
&=& 1 - \Big[0.023\,(\mbox{NLO}) + 0.030\,(\mbox{NNLO})\Big]\,\,(R_{\perp})\, 
\nonumber\\
&& + 0.084\,\Big\{1+0.406\,(\mbox{NLO spec.})
+ 0.032 \,(\delta_{\rm log}^\parallel)\Big\}
\nonumber\\[0.2cm]
&=& 1.067.
\end{eqnarray}
In these expressions we separated the symmetry-conserving~(first
number, normalized to 1), A0- and B-type corrections (denoted by
$R_{T,\perp}$ and the remaining terms, respectively).
The parameter $\delta_{\rm log}^\parallel$ denotes the small effect
from renormalization-group summation and has the same meaning as in
Eq.~(124) of~\cite{Beneke:2005gs}. We observe that 
the A0-type and spectator-scattering corrections always have opposite 
sign, but the latter are larger and determine the sign of the 
deviation from the symmetry limit. We also notice that
the two-loop correction to $R_{\perp}$ is more significant than to
$R_{T}$. The small numerical difference of spectator-scattering
contribution relative to Eq.~(124) in \cite{Beneke:2005gs} is due to
the fact that now the 3-loop running coupling is used. For comparison
the QCD sum rule calculation~\cite{Ball:2004ye} gives ${\cal
  R}_1=0.955$ and ${\cal R}_2=0.947$. For the tensor-to-vector ratio
${\cal R}_2$, one notices that the sign of the symmetry-breaking
correction between these two methods is opposite. Since the 
form factor ratio $T_1/V$ is important for radiative and electroweak 
penguin decays 
(see the discussion in Section~5.2 of \cite{Beneke:2005gs}), 
the discrepancy between the SCET and QCD sum rules results for 
${\cal R}_2$ suggests that a dedicated analysis of symmetry 
breaking corrections to form factors (rather than the form factors 
themselves) with the QCD sum rule method should be performed.

\section{Semi-inclusive $\bar B\to X_s\ell^+\ell^-$ decays}
\label{sec:inclusive}

Rare inclusive $B$-meson decays induced by the quark level transition 
$b\to s \ell^+ \ell^-$ are highly sensitive to new physics. Due to the
presence of two extra operators $(\bar\ell\ell)_{V,A} (\bar s
b)_{V-A}$ in the effective Hamiltonian and the availability of 
additional kinematical observables, such as the dilepton invariant 
mass~($q^2$) spectrum and the forward-backward asymmetry, the
$b\to s \ell^+ \ell^-$ decay provides complementary information 
relative to the radiative $b\to s \gamma$ process.

The exclusive decay process  $B\to K^{*} \ell^+\ell^-$ has been 
studied in great detail, both with respect to its 
QCD dynamics~\cite{Beneke:2001at} and to the 
sensitivity of various observables to new physics
\cite{Bobeth:2008ij}, because it can be measured relatively easily 
at hadron colliders. Also on the inclusive decay process
$\bar B\to X_s \ell^+\ell^-$ dedicated work exists on higher order
radiative corrections~(see~\cite{Hurth:2007xa} for recent reviews),
power corrections~\cite{Falk:1993dh,Buchalla:1997ky},
and on the identification of additional kinematic
observables~\cite{Lee:2006gs}.

The low dilepton invariant mass region, 
$1\,{\rm GeV^2}\leq q^2 \leq 6\,{\rm GeV^2}$ is particularly 
interesting, since it benefits from smaller theoretical uncertainties 
and a higher rate. At somewhat higher $q^2$ the spectrum is dominated 
by charmonium resonances (which also determine the integrated 
decay rate, see the discussion in \cite{Beneke:2009az}). On the other
hand, for $q^2 < 1\,{\rm GeV^2}$, the branching ratio is determined 
largely by the contribution from almost real intermediate photons, and
hence contains essentially the same information as the 
$b \to s \gamma$ transition.

In the following we discuss semi-inclusive $\bar B\to X_s \ell^+\ell^-$
decay, where the hadronic final state $X_s$ is constrained to have 
small invariant mass $m_X$ and $q^2$ is in the range from 
$1\,{\rm GeV^2}$ to $6\,{\rm GeV^2}$. In this kinematic region~(the 
so-called ``shape function region''), the outgoing hadronic state is 
jet-like and the relevant degrees of freedom are hard-collinear and 
soft modes. The semi-inclusive decay rates can be calculated by 
matching the effective weak interaction Hamiltonian to 
soft-collinear effective theory. At the leading order in the 
$\lqcd/m_b$ expansion, the decay rates can be factorized into  
process-dependent hard functions $h^{[0]}$, related to physics at the 
hard scale $\mu\sim m_b$ and above, a universal jet function $J$, 
related to physics at the intermediate hard-collinear scale 
$\mu_{\rm hc}\sim \sqrt{m_b\lqcd}$, as well as a universal non-perturbative 
shape function $S$, describing the internal soft dynamics of the 
$B$ meson, with the following schematic form~\cite{Lee:2005pk,Lee:2008xc}
\begin{equation}
\label{eq:fact}
\mathrm{d} \Gamma^{[0]} = h^{[0]} \times J \otimes S\,,
\end{equation}
a result already applied extensively to inclusive 
$\bar B\to X_u\ell\bar\nu$ and $\bar B\to X_s\gamma$ decays in the
shape-function region. The two-loop matching coefficients of the
tensor currents calculated in the present paper provide further 
input to reaching NNLO ($\alpha_s^2$) accuracy in $h^{[0]}$ 
and the entire differential decay rate $\mathrm{d} \Gamma^{[0]}$. 
Compared to exclusive decays mediated by the $b\to s\ell^+\ell^-$ 
transition~\cite{Beneke:2001at} the semi-inclusive case has the 
advantage that the theoretically less certain spectator-scattering 
contributions to the currents that enter the exclusive form 
factors are power-suppressed and can be dropped.

In the following we will be mainly interested in the forward-backward 
asymmetry of the differential rate integrated up to an invariant 
mass $m_X^{\rm cut}$ in the final state. We briefly review the theoretical 
description of this quantity, adopting the same conventions and 
notation as \cite{Lee:2008xc}, to which we also refer for further 
details. The short-distance coefficients $h^{[0]}$ at the 
hard matching scale $\mu$ are composed of products of two factors, 
since the hadronic part of the effective weak interaction Hamiltonian 
is first matched to two QCD (rather than SCET) currents,
\begin{equation}\label{eq:effcurrents}
J_9^\mu = \bar{s}\,\gamma^\mu P_L b \,,\qquad
J_7^\mu = \frac{2\,m_b}{q^2}\, 
\bar{s}\, iq_\rho\sigma^{\rho\mu} P_R b\, \Big\vert_{\nu=m_b} \,,
\end{equation}
with coefficients $C_i^{\rm incl}(q^2,\mu)$ and $P_{L,R}=(1\mp\gamma_5)/2$.
Moreover, $m_b$ in $J_7^\mu$ refers to the bottom quark pole mass.
The QCD currents are then related to the corresponding SCET currents,
\begin{eqnarray}\label{eq:SCETmatching}
J_9^\mu &=& \sum_{i=1,2,3} c^9_i(u,\mu)\,
[\bar{\xi}W_{hc}]\,\Gamma_{9,i}^\mu\,h_v\,,\nonumber \\
J_7^\mu &=& \frac{2m_b}{q^2}\,\sum_{i=1,2} c^7_i(u,\mu)\,
[\bar{\xi}W_{hc}]\,\Gamma^\mu_{7,i}\,h_v\,.
\end{eqnarray}
These equations represent the
momentum space versions of (\ref{eq:matching}). The variable $u$ is 
related to the kinematics of the process by $u=p^-/m_b$, 
where
\begin{equation}
p^-= n_+ p = m_b-\frac{q^2}{m_B-p_X^+},
\label{defpm}
\end{equation}
and $p_X^+ = n_- p_X \ll m_B$ is the small light-cone component of 
the hadronic final state's momentum. The basis of Dirac structures is 
chosen as
\begin{eqnarray}
\label{DiracstructureNew}
\Gamma_{9,i}^\mu &=& P_R\,\bigl\{\gamma^\mu, v^\mu, q^\mu \bigr\}\,,
\nonumber \\
\Gamma_{7,i}^\mu &=& P_R\,\bigl\{i q_\nu\sigma^{\nu\mu}, 
q_\nu (q^\nu v^\mu-q^\mu v^\nu) \bigr\}\,.
\end{eqnarray}
As noted in \cite{Lee:2008xc}, the choice of $q^\mu$ instead of
$n_{-}^\mu$ for $\Gamma^\mu_{9,3}$ is convenient here as it makes 
explicit the constraint from lepton current conservation, which
implies that for massless leptons $c^9_3$ does not contribute, 
while for $\Gamma^\mu_{7,i}$ there are only two independent
coefficients. Transforming the basis (\ref{DiracstructureNew}) to our 
operator basis listed in Table~\ref{tab:basis}, the matching
coefficients $c^9_i$ and $c^7_i$ are given, respectively, as
\begin{eqnarray}
&& c^9_1(u,\mu) = C^1_V(u;\mu)\,, \qquad \nonumber \\
&& c^9_2(u,\mu) = C^2_V(u;\mu)+\frac{2}{u}\,C^3_V(u;\mu)\,, \qquad  \nonumber \\
&& c^9_3(u,\mu) = -\frac{2}{u m_b}\,C^3_V(u;\mu)\,, \nonumber \\
&& c^7_1(u,\mu) = -2\,C^1_T(u;\mu,\nu=m_b)+C^3_T(u;\mu,\nu=m_b)\,, \qquad  \nonumber \\
&& c^7_2(u,\mu) =-\frac{2}{u m_b}\,C^3_T(u;\mu,\nu=m_b)\,.
\end{eqnarray}
The two-loop matching coefficients $c^9_i$ for the vector current have 
become available in the context of inclusive semi-leptonic $B$
decays~\cite{Bonciani:2008wf,Asatrian:2008uk,Beneke:2008ei,Bell:2008ws}.
The results of this paper allow us to compute also the matching
coefficients $c^7_i$ at NNLO. As a consequence the factor in 
$h^{[0]}$ related to the QCD current matching is now complete at NNLO, while 
the other factor related to $C_i^{\rm incl}(q^2,\mu)$ is 
known at the next-to-next-to-leading logarithmic (NNLL) order, since 
the three-loop ${\cal O}(\alpha_s^2)$ matrix elements of 
the current-current operators (giving rise to charm-loop diagrams) 
are not available.

\begin{figure}[t!]
\begin{center}
\epsfig{file=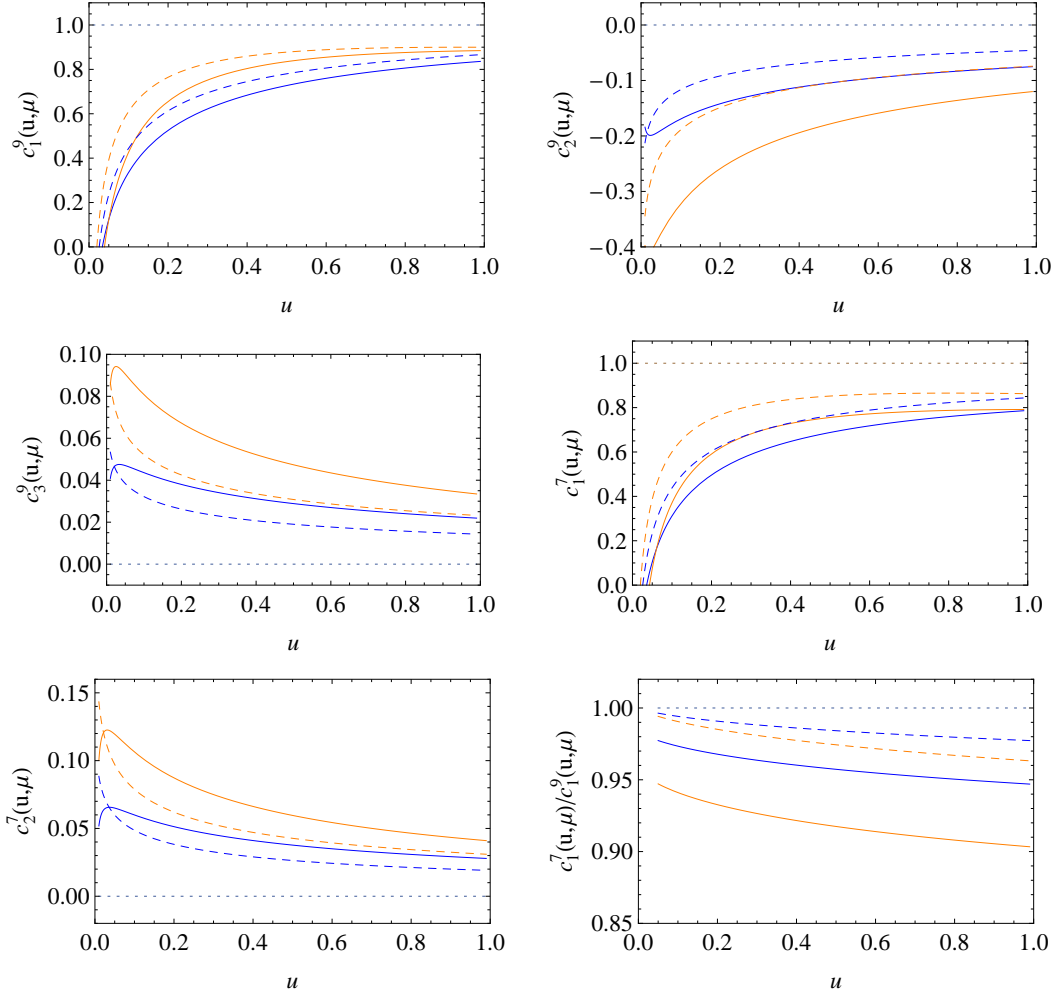,width=14cm}
\caption{\label{fig:matching} \small \textit{The matching 
coefficients $c^9_i(u,\mu)$ and $c^7_i(u,\mu)$ as a function 
of $u$~(related to the dilepton invariant mass $q^2=(1-u)m_b^2$) 
in the one-loop (dashed) and two-loop (solid) approximation.
The blue/dark grey curves refer to $\mu=m_b=4.8\,$GeV, 
and the orange/light grey ones to $\mu=1.5\,$GeV. 
}}
\end{center}
\end{figure}

In Figure~\ref{fig:matching} we show these matching coefficients as a 
function of $u$ in the one- (dashed) and two-loop (solid)
approximation, evaluated at $\mu=m_b=4.8\,$GeV (blue/dark grey curves)
and at $\mu=1.5\,$GeV (orange/light grey curves), respectively.
The difference between these two different choices of
the IR factorization scale $\mu$ is compensated by the 
corresponding scale dependence of the convolution $J \otimes S$ such
that the differential rate (\ref{eq:fact}) is $\mu$-independent. 
Note that, while we show the entire range of $u$, Eq.~(\ref{defpm}) 
implies that the relevant values of $u$ for $b\to s \ell^+\ell^-$ in 
the $q^2$ region of interest are above $u\approx 0.75$. In the lower 
right panel of Figure~\ref{fig:matching}, we also show the ratio
$c_1^7/c_1^9$,
which equals the quantity $R_\perp$ defined earlier in
(\ref{rfactors}) at $\nu=m_b$, and plays an important role for the forward-backward
asymmetry as discussed below. Note that $R_\perp$ is
$\mu$-independent, except for the truncation 
of the perturbative series. In evaluating this ratio to a given 
order in $\alpha_s$, we expand the denominator and truncate the 
expanded expression.

Comparing the dashed (one-loop approximation) and solid (two-loop 
approximation) curves of the same colour in
Figure~\ref{fig:matching}, we observe that the two-loop corrections
are generally moderate in the large $u$~(low $q^2$) region,
whereas the large correction in the region of small $u$ is due to the
fact that increasing powers of large logarithms take over in this
region. However, the correction is amplified in the ratio 
$R_\perp$, where the two-loop correction exceeds the one-loop term. 
This leads to a considerable residual $\mu$-dependence (difference 
of blue/dark grey and orange/light grey curves) as can also be seen 
in Figure~\ref{fig:scale_dep}. Since the infrared physics drops 
out from the ratio $c_1^7/c_1^9$ the natural scale is of 
order of the hard scale $m_b$.

The differential decay rate (\ref{eq:fact}) can be written as 
\begin{eqnarray}
\frac{d^3\Gamma}{d q^2 dp_X^+ d\cos\theta} &=& \frac{3}{8} \,\Big[
(1+\cos^2\theta) H_T(q^2,p_X^+) + 
2\,(1-\cos^2\theta) H_L(q^2,p_X^+) 
\nonumber\\ 
&& +\,2\cos\theta  \,H_A(q^2,p_X^+)\Big],
\end{eqnarray}
where for $\bar B$ decay, $\theta$ denotes the angle between the
positively charged lepton and the $\bar B$ meson in the 
centre-of-mass frame of the $\ell^+\ell^-$ pair. For fixed $p_X^+$, the 
forward-backward asymmetry in $\theta$ therefore vanishes for
a particular $q_0^2$ 
at which $H_A(q_0^2,p_X^+)=0$. Integrating over the invariant mass 
of the hadronic final state up to the cut $m_X^{\rm cut}$, the 
asymmetry zero occurs at 
\begin{eqnarray}
0 &=& \int_0^{p_X^{+\rm cut}} \hspace*{-0.3cm} 
dp_X^+ \,H_A(q_0^2,p_X^+) 
\nonumber\\
&=& {\rm const} \times 
\int_0^{p_X^{+\rm cut}}  \hspace*{-0.3cm} 
dp_X^+ \, h_A^{[0]}(q_0^2,p_X^+) \,
\frac{(q_{0+}-q_{0-})^2}{q_{0+}}\,q_0^2
\int d\omega \,p^- J(p^-\omega)\,S(p_X^+-\omega),
\label{FBasymcondition}
\end{eqnarray}
where \cite{Lee:2008xc} $q_+=m_B-p_X^+$, $q_-=q^2/q_+$,
\begin{equation}
p_X^{+\rm cut} = 
\frac{1}{2 m_B}\left[
m_B^2+({m_X^{\rm cut}})^2-q^2 - 
\sqrt{(m_B^2+({m_X^{\rm cut}})^2-q^2)^2 - 4 m_B^2 ({m_X^{\rm cut}})^2 }
\,\right],
\end{equation}
and 
\begin{equation}
h_A^{[0]}(q^2,p_X^+) = 2 {\cal C}_{10} \,c_1^9(u) \,
\mbox{Re}\left[C_9^{\rm incl}(q^2)  c_1^9(u) + 
\frac{2 m_b}{q_-}\,C_7^{\rm incl}(q^2)  c_1^7(u)
\right]\,.
\label{hA}
\end{equation}

We now observe that $h_A^{[0]}(q_0^2,p_X^+)$ depends on
$p_X^+$ only through the definition of $u$ in (\ref{defpm}) and the
kinematic factor $2m_b/q_-$. For typical $m_X^{\rm cut}$ of $2\,$GeV
this dependence is very weak, since then $p_X^+ \sim 1\,\mbox{GeV}\ll 
m_B$. Thus, $p_X^+$ appears only as a small correction to $m_B-p_X^+$,
and in the definition of $u$ in a term
that is additionally suppressed by $q^2/m_B$ relative to $m_b$,
see (\ref{defpm}).
This results in a very small variation of $u$ of about 0.02 
over the entire $p_X^+$ integration region. We may therefore pull the 
slowly varying function  $h_A^{[0]}(q_0^2,p_X^+)$ in front 
of the $p_X^+$ integration in (\ref{FBasymcondition}) thereby 
replacing $p_X^+$ in the argument by an average value which we assume 
to be $\langle p_X^+\rangle = p_X^{+\rm cut}/2$. The remaining 
integral over the jet and soft function is different from zero, thus 
the forward-backward asymmetry zero is determined by 
$h_A^{[0]}(q_0^2,\langle p_X^+ \rangle) =0$. Using (\ref{hA}) 
this is equivalent to the condition 
\begin{equation}
\frac{q_0^2}{2 m_b (m_B-\langle p_X^+ \rangle)} 
= - \, \frac{\mbox{Re}\,[C_7^{\rm incl}(q_0^2)]}
{\mbox{Re}\,[C_9^{\rm incl}(q_0^2)]}\,
\frac{c_1^7(u_0)}{c_1^9(u_0)}
\label{q0eq}
\end{equation}
with $u_0\equiv 1-q_0^2/(m_b(m_B-\langle p_X^+ \rangle))$. This result 
leads to the important conclusion that the 
{\em QCD dynamics that determines the location of the 
asymmetry zero is to a very good approximation
independent of the long-distance physics 
below the scale $m_b$ contained in the jet function and the
non-perturbative shape function.} It also depends only very weakly on 
the value of the invariant mass cut through the dependence 
of $\langle p_X^+ \rangle$ on $m_X^{\rm cut}$.
The bulk dependence of $q_0^2$ on the
invariant mass cut $m_X^{\rm cut}$ enters through the kinematical
factor $m_B-\langle p_X^+\rangle$ on the left-hand side of~(\ref{q0eq}).

We are now in the position to quantify the impact of the two-loop 
calculation of $R_\perp(u_0,\nu=m_b) = c_1^7(u_0)/c_1^9(u_0)$ on $q_0^2$. 
In \cite{Lee:2008xc} the asymmetry zero has been determined 
by keeping the full NNLL expression for 
$\mbox{Re}\,[C_7^{\rm incl}(q^2)]/\mbox{Re}\,[C_9^{\rm incl}(q^2)]$ but 
setting $R_\perp=1$. In this approximation, and excluding
$1/m_b$-suppressed shape function effects for the moment,
the zero is found to be
\begin{equation}
q_0^2{}_{\big|_{R_\perp=1}} = (3.62 \, \ldots \, 3.69) \, \mbox{GeV}^2 \qquad \mbox{for} \qquad m_X^{\rm cut} 
= (2.0 \ldots 1.8)\,\mbox{GeV} \, .\label{eq:zeroRperpunity}
\end{equation}
As indicated the lowest value corresponds to $m_X^{\rm cut} =2.0$~GeV
and the highest one to $m_X^{\rm cut} =1.8$~GeV.
Our value is somewhat larger than what can be extracted from 
Figure~4 of~\cite{Lee:2008xc}, because we expand the factor 
$\overline{m}_b(\mu)/m_b^{\rm pole}$ that accompanies $C_7$
in $\alpha_s$. Moreover, the variation of the zero when 
changing $m_X^{\rm cut}$ from $1.8$~GeV to $2.0$~GeV is about twice 
as large compared to what can be read off
from Fig.~4 of~\cite{Lee:2008xc}, which is likely due to our approximation of 
pulling the slowly varying function $h_A^{[0]}(q_0^2,p_X^+)$ out of 
the integral in (\ref{FBasymcondition}). However, our approximation 
is still justified since
even the increased sensitivity of the zero on $m_X^{\rm cut}$ is only
$\pm 0.03$~GeV$^2$ and hence below~$1$\%. 
Taking into account $R_\perp$ at the NLO, we find for the position of the zero
\begin{equation}
q_0^2{}_{\big|_{R_\perp \, {\rm NLO}}} = (3.55 \, \ldots \, 3.61) \, \mbox{GeV}^2 \qquad \mbox{for} \qquad m_X^{\rm cut} 
= (2.0 \ldots 1.8)\,\mbox{GeV} \, . \label{eq:zeroRperpNLO}
\end{equation}
The impact of the NLO correction to $R_\perp$ is to shift the zero by $-2.2$\%.
As we already stated before, and as can also be seen from
Figures~\ref{fig:scale_dep} and~\ref{fig:matching}, the size of the 
NNLO correction to $R_\perp$ is significant. It amounts to a shift of the NLO zero
in~(\ref{eq:zeroRperpNLO}) by another $-3$\% and hence is larger than the NLO shift.
The total shift induced by $R_\perp$ through NNLO therefore amounts 
to $-5$\%.
\begin{table}[t]
\begin{center}
\begin{displaymath}
\begin{tabular}{|l|l|}
\hline
\spp $\alpha_s (M_Z) = 0.1180 $ & $\lambda_2 \simeq \frac{1}{4} \left(m_{B^*}^2-m_B^2\right) \simeq 0.12 \;{\rm GeV}^2$ \\[-0.3cm]
\spp $\sin^2\theta_W = 0.23122$ & $m_{t}^{{\rm pole}}= 171.4  \; {\rm GeV} $
\\[-0.3cm]
\spp $M_W = 80.426\;{\rm GeV}$  &  $m_c^{\rm pole}= (1.5 \pm 0.1)
\;{\rm GeV}$   \\[-0.3cm]
\spp $M_Z = 91.1876\;{\rm GeV}$ & $m_b^{\rm PS}(2\, {\rm GeV}) = (4.6 \pm 0.1)\;{\rm GeV}$ \\\hline
\end{tabular}
\end{displaymath}
\caption{\small
\textit{Numerical inputs that we use in the phenomenological 
analysis of the forward-backward asymmetry zero.}}
\label{tab:inputs} 
\end{center}
\end{table}

Before proceeding to our final result we briefly comment on the r\^ole of
power corrections. 
The authors of \cite{Lee:2008xc} performed a thorough study of
$1/m_b$-suppressed shape function effects which result in a shift
of the zero of $-0.05\,$GeV$^2$ to $-0.1\,$GeV$^2$.
This shift is more strongly 
dependent on the invariant mass cut and the theoretical error 
increases when $m_X^{\rm cut}$ is chosen smaller.
In the following we take the larger value as an estimate for the shift
and also for the associated uncertainty. However, the study of power corrections 
in \cite{Lee:2008xc} does not cover all such corrections and applies 
a rather crude treatment to those arising from soft gluon attachments 
to the charm-loop diagrams by absorbing the $1/m_c^2$ non-perturbative
power corrections into the $C_i^{\rm incl}$, which is justified 
only in the absence of invariant mass cuts. In the semi-inclusive
region, the matrix element of~(29) in~\cite{Buchalla:1997ky} cannot, 
due to the presence of a soft gluon, be expressed in terms of 
a short-distance coefficient times a local matrix element, since 
the soft gluon attached to the charm loop affects
the invariant mass of an energetic hadronic final state by a relevant 
amount $\sqrt{m_b \Lambda_{\rm QCD}}$, which must be accounted for by a
subleading shape function. By treating this correction as in the 
inclusive case, the authors of~\cite{Lee:2008xc} implicitly
assumed that this shape function somehow factorizes into the
local heavy-quark effective theory  
matrix element $\lambda_2$ and the leading-power shape function. 
It is not clear to us how this simplification can be justified 
and it is likely not even parametrically correct. Nevertheless, 
in the absence of better information we follow the treatment 
of~\cite{Lee:2008xc} and include the  $1/m_c^2$ 
power corrections into the $C_i^{\rm incl}$. This results in 
a shift of the asymmetry zero by $+0.07\,$GeV$^2$, which is
included in~(\ref{eq:zeroRperpunity}),~(\ref{eq:zeroRperpNLO}),
and below in~(\ref{eq:zeroRperpNNLO}).
To be conservative we assign 
another $0.1\,$GeV$^2$ uncertainty to this estimate and add it in quadrature 
with the other power correction uncertainty.
 
We are now in the position to present our final NNLO result based on the 
numerical input parameters and their respective intervals as specified in Table~\ref{tab:inputs}. 
We then find
\begin{eqnarray}
q_0^{\,2} &=& \big[(3.34 \, \ldots \, 3.40) \, {}^{+0.04}_{-0.13} {}_{\, \mu} 
\pm 0.08{}_{\,m_b} \, {}^{+0.05}_{-0.04} {}_{\, \,m_c}
\pm {0.14}_{\,{\rm SF}}
\pm 0.14 {}_{\,\langle \, p_X^+\rangle}\big]\,\mbox{GeV}^2  
\nonumber \\
&=& \big[(3.34 \, \ldots \, 3.40) {}^{+0.22}_{-0.25}\big]\,\mbox{GeV}^2 
\qquad \mbox{for} \qquad m_X^{\rm cut} 
= (2.0 \ldots 1.8)\,\mbox{GeV} \, . \label{eq:zeroRperpNNLO}
\end{eqnarray}
The error estimate is computed as follows:
The range of scale variation is taken to be 
$2.3$~GeV~$<\mu<$~$9.2$~GeV, and we vary the 
scale in the $C_i^{\rm incl}$ and in $R_\perp$ independently to 
account conservatively for the absence of the ${\cal
  O}(\alpha_s^2)$ correction to the $C_i^{\rm incl}$. 
The input quark mass is the bottom mass in the 
potential-subtracted (PS) scheme~\cite{Beneke:1998rk}, see 
Table~\ref{tab:inputs}. The pole mass and $\overline{\rm MS}$ 
mass used in intermediate expressions are computed using the 
one-loop conversion factors resulting in  $m_b^{\rm pole} = 
4.78\,$GeV and $\overline{m}\,(m_b^{\rm PS}) = 4.36\,$GeV, respectively,
when $m_b^{\rm PS}(2\, {\rm GeV}) = 4.6\,$GeV. The dependence 
on the charm quark mass enters through the matrix elements 
of the current-current operators. The error labelled ``SF'' 
is connected with the subleading shape function effects as 
discussed above. Finally we have added an uncertainty estimate 
for the approximation made by pulling out the 
slowly varying function $h_A^{[0]}(q_0^2,p_X^+)$ out of 
the $p_X^+$ integral in (\ref{FBasymcondition}). We estimate 
this error rather generously by varying $\langle p_X^+\rangle$ 
from $p_X^{\rm cut}/4$ to  $3 p_X^{\rm cut}/4$. The total error is 
obtained by adding all these uncertainties in quadrature.

We note that the value of the 
asymmetry zero in semi-inclusive $b\to s \ell^+\ell^-$ 
decay is significantly smaller than for the exclusive 
case~\cite{Beneke:2001at}, where spectator scattering is responsible for a
positive shift as is the fact that in this case 
$\langle p_X^+\rangle =0$ in (\ref{q0eq}).
 On the other hand the semi-inclusive zero is in the same region as
in the inclusive case~\cite{Huber:2007vv}, where
virtual effects together with hard gluon bremsstrahlung
encoded in functions $\omega_{710}$ and $\omega_{910}$~\cite{Asatrian:2002va} also induce
a negative shift on the zero.
%
%

\section{Conclusion}
\label{sec:conclusion}


In this paper we completed the two-loop matching calculation for 
heavy-to-light currents from QCD onto SCET for the complete set of Dirac 
structures. These matching coefficients enter several phenomenological 
applications, of which we have discussed their effects on heavy-to-light 
form factor ratios, exclusive radiative and semi-leptonic decays, as well 
as the inclusive decay $\bar B \to X_s \ell^+\ell^-$ in the shape-function 
region. The two-loop corrections are generally relatively small, in the 
few percent range. However, one ratio, $R_{\perp}= c^7_1(u,\mu)/c^9_1(u,\mu)$, 
which is also the most important for phenomenology, since it enters 
the comparison of radiative and semi-leptonic decays as well as the 
forward-backward asymmetry in exclusive and semi-inclusive 
$b\to s\ell^+\ell^-$ transition, exhibits a two-loop correction  
that is larger than the one-loop term. The two-loop term alone shifts 
the location of the asymmetry zero by about $-0.1\,$GeV$^2$, comparable 
to the effect of $1/m_b$ suppressed shape functions estimated 
in~\cite{Lee:2008xc}. We showed that the location of the 
asymmetry zero in semi-inclusive $\bar B \to X_s \ell^+\ell^-$ with 
an invariant mass cut is to a very good approximation
independent of the long-distance physics 
below the scale $m_b$ contained in the jet function 
and the non-perturbative shape function, and obtain 
$q_0^{\,2} = (3.34 {}^{+0.22}_{-0.25})\,\mbox{GeV}^2$
for an invariant mass cut $m_X^{\rm cut}=2.0\,$GeV as our best estimate 
for the asymmetry zero. Moreover, we confirm the discrepancy between 
QCD sum rule and SCET results for the form factor ratio 
$T_1/V$ in the low $q^2$ region discussed in \cite{Beneke:2005gs} and 
suggest that a dedicated QCD sum rules analysis of deviations 
from the symmetry limit (rather than the form factors themselves) 
should be done to clarify the situation.


\subsubsection*{Acknowledgements}
We would like to thank F.~Tackmann and M.~Misiak for useful correspondence.
This work was supported in part by the DFG Sonder\-forschungsbereich/Transregio~9
``Computergest\"utzte Theoretische Teilchenphysik'' (GB, MB), 
the Helmholtz alliance ``Physics at the Terascale'' (TH), and  
the Alexander-von-Humboldt Stiftung (X.-Q.~Li).
X.-Q.~Li acknowledges hospitality from the Institute of Theoretical Physics, 
Chinese Academy of Science, where part of this work was performed.


\begin{appendix}

\section{NLO coefficient functions}
\label{app:1loop}

In Section~\ref{sec:Fij} we introduced the following set of one-loop
coefficient functions,
\begin{align}
g_0(u) &= - \frac52 + 2 \ln (u),
\no\\
g_1(u) &=  - \frac{\pi^2}{12} + \frac{2}{\ub} \ln (u) - 2 \ln^2(u) - 2
\Li_2(\ub),
\no\\
g_2(u) &= \frac{\pi^2}{24} + \frac13 \zeta_3 +
\frac{12(1+\ub)+\pi^2\ub}{6\ub} \ln(u) - \frac{2}{\ub}
\Big( \ln^2(u) + \Li_2(\ub) \Big) + \frac43 \ln^3(u)
\no\\
&\quad
 + 4 \ln(u) \Li_2(\ub) - 2 \Li_3(\ub) + 4
\S_{1,2}(\ub),
\no\\
g_3(u) &= - \frac{\pi^4}{160} - \frac16 \zeta_3 +
\frac{48(1+\ub)+2\pi^2-8\ub \zeta_3}{12\ub} \ln(u)
- \frac23 \ln^4(u) - 4 \ln^2(u) \Li_2(\ub)
\no\\
&\quad
- \frac{12(1+\ub)+\pi^2\ub}{6\ub} \Big( \ln^2(u) + \Li_2(\ub) \Big)
- 8 \ln(u) \S_{1,2}(\ub) + 4 \ln(u) \Li_3(\ub)  - 2 \Li_4(\ub)
\no\\
&\quad
+ \frac{2}{\ub} \bigg( \frac23 \ln^3(u) + 2 \ln(u) \Li_2(\ub) -
\Li_3(\ub) + 2 \S_{1,2}(\ub) \bigg)  - 8 \S_{1,3}(\ub) + 4
\S_{2,2}(\ub),
\no\\
g_4(u) &= g_1(u) - 6 - \frac{4u}{\ub} \ln(u),
\no\\
g_5(u) &= g_2(u) - 10 - \frac{\pi^2}{3} - \frac{2u}{\ub} \Big( 3\ln(u) - 2\ln^2(u) -
2\Li_2(\ub) \Big),
\no\\
g_6(u) &= g_3(u) - 18 - \frac{\pi^2}{2} + \frac43 \zeta_3 - \frac{2u}{\ub} \bigg(
\frac{30+\pi^2}{6} \ln(u) - 3\ln^2(u) - 3\Li_2(\ub) + \frac43 \ln^3(u)
\no\\
&\quad
+ 4 \ln(u) \Li_2(\ub) - 2\Li_3(\ub) + 4 \S_{1,2}(\ub) \bigg),
\no\\
g_7(u) &= - \frac{2}{\ub} - \frac{2u}{\ub^2} \ln(u),
\no\\
g_8(u) &= - \frac{6}{\ub} - \frac{2u}{\ub^2} \Big( 2\ln(u) - \ln^2(u) -
\Li_2(\ub) \Big),
\no\\
g_{9}(u) &= \frac{2u}{\ub} \ln(u),
\no\\
g_{10}(u) &= \frac{u(1+4\ub)}{\ub^2} \ln(u) + \frac{u}{\ub}
\Big( 1 - 2 \ln^2(u) - 2 \Li_2(\ub) \Big),
\no\\
g_{11}(u) &= \frac{6u(2+7\ub)+\pi^2 u \ub}{6\ub^2} \ln(u) -
\frac{u(1+4\ub)}{\ub^2} \Big(\ln^2(u)+\Li_2(\ub)\Big)
\no\\
&\quad
+ \frac{u}{\ub} \bigg( 3 + \frac43 \ln^3(u) + 4 \ln(u) \Li_2(\ub) -
2 \Li_3(\ub) + 4\S_{1,2}(\ub) \bigg),
\no\\
g_{12}(u) &= g_7(u) + 2,
\no\\
g_{13}(u) &= g_8(u) + 6 + \frac{2u}{\ub} \ln(u).
\end{align}

\section{NNLO coefficient functions}
\label{app:2loop}

The finite parts of the two-loop form factors involve the following 
coefficient functions,
\begin{align}
h_1(u) &=
-\frac{2 (7 - 2 \ub + 3 \ub^2)}{u^2} \Li_4(\ub)
- \frac{4(11 + 2 \ub + 3 \ub^2)}{u^2} \S_{2,2}(\ub)
+ 8 \S_{1,3}(\ub)
- 8 \ln(u) \Li_3(\ub)
\no\\
&
+ \frac{2 (3 + \ub^2)}{u^2} \Li_2(\ub)^2
+ 16 \ln(u)\S_{1, 2}(\ub)
+ \frac{16}{3} \ln^4(u)
+ 16 \ln^2(u)\Li_2(\ub)
\no\\
&
- \frac{6 + 47 \ub - 5 \ub^2}{3 u \ub} \Li_3(\ub)
- \frac{2 (42 - 29 \ub)}{9 \ub}\ln^3(u)
+  \frac{2(6 - 115 \ub + 13 \ub^2)}{3 u\ub} \S_{1, 2}(\ub)
\no\\
&
- \frac{2(12 + \ub + 11 \ub^2)}{3 u \ub} \ln(u) \Li_2(\ub)
+ \frac{36 - 87 \ub - 250 \ub^2 + 18 \pi^2 \ub^2}{9 \ub^2} \ln^2(u)
\no\\
&
- \bigg( \frac{33 + 109 \ub - 322 \ub^2}{9 u \ub}
 - \frac{(7 - 2 \ub + 3 \ub^2) \pi^2}{u^2}\bigg)\Li_2(\ub)
+ \frac{(2815 + 353 \ub)\pi^2}{432 u}
\no\\
&
+ \bigg( \frac{1173 + 241 \ub}{27\ub} - \frac{(9 + 46 \ub + 17 \ub^2)\pi^2}{9u\ub}
 -\frac{56}{3} \zeta_3 \bigg)  \ln(u)
\no\\
&
- \frac{(509 + 278 \ub + 77 \ub^2) \pi^4}{720 u^2}
+ \frac{76}{9} \zeta_3 + \frac{30331}{1296} - 2 h_2(u),
\no\\
h_2(u) &=
\frac{2 (1 + \ub)^2}{3u\ub} \Big( 12\calH_1(\ub) + \pi^2 \ln(2-u)\Big)
+ \frac13 \Big( 24 \calH_2(\ub)- 2\pi^2 \Li_2(-\ub)\Big)
\no\\
&
- \frac{8}{u^2} \S_{2,2}(\ub)
- \frac{2}{u^2} \Li_4(\ub)
- 8 \ln(u)\Li_3(\ub)
- \frac{(u - \ub) (3 - 2 \ub)}{u^2}  \Li_2(\ub)^2
\no\\
&
- \frac{40 - 56 \ub + 7 \ub^2}{3u^2} \Li_3(\ub)
+ \frac{14 - 40 \ub + 17 \ub^2}{3u^2} \S_{1,2}(\ub)
+ \frac{29 - 35 \ub}{3u} \ln(u)\Li_2(\ub)
\no\\
&
+ \frac{44}{9} \ln^3(u)
- \bigg( \frac{66 + 122 \ub - 89 \ub^2}{9 u \ub}
- \frac{(7 - 8 \ub + 4 \ub^2) \pi^2}{3 u^2} \bigg) \Li_2(\ub)
\no\\
&
- \frac{78 + 223 \ub - 12 \pi^2 \ub}{18\ub} \ln^2(u)
+ \frac{(13 - 62 \ub + 31 \ub^2) \pi^4}{120u^2}
\no\\
&
+ \bigg( \frac{2 (354 + 121 \ub)}{27\ub} - \frac{(24 - 71 \ub + 65 \ub^2) \pi^2}{18u\ub}
- 14 \zeta_3 \bigg) \ln(u)
\no\\
&
+ \frac{3 (2 - \ub)^2}{u^2}  \bigg( \Li_3(-u) - \ln(u)\Li_2(-u) -
\frac{\ln^2(u)+\pi^2}{2}\ln(1+u)\bigg)
+ \frac{5405}{1296}
\no\\
&
+ \frac{(877 + 239 \ub) \pi^2}{216u}
+ \frac{469 - 73 \ub}{18 u}  \zeta_3
- \frac{2 (3 + \ub) \pi^2}{u} \ln(2),
\no\\
h_3(u) &=
\frac83 \Li_3(\ub)
+ \frac{8 (1 + \ub) (3 + 11 \ub - 11 \ub^2 + 5 \ub^3)}{9 u^3 \ub} \Li_2(\ub)
\no\\
&
- \frac{2 (96 + 208 \ub - 224 \ub^2 + 112 \ub^3 + 3 \ub u^2 \pi^2)}{27 u^2 \ub}  \ln(u)
+ \frac{ 3773 - 4954 \ub + 2333 \ub^2}{81u^2}
\no\\
&
- \frac{(265 - 315 \ub + 219 \ub^2 - 41 \ub^3) \pi^2}{54 u^3}
- \frac{28}{9} \zeta_3,
\no\\
h_4(u) &=
h_1(u)  + 2 h_2(u)
+ \frac{12 (1 + \ub)^2}{u^3}
\bigg( 8 \S_{2,2}(\ub) + 2 \Li_4(\ub)  - \Li_2(\ub)^2 + \frac{3 \pi^4}{20} \bigg)
\no\\
&
+ \frac{4 (1 + \ub) (1 + 10 \ub + \ub^2)}{u^2\ub} \Li_3(\ub)
+ \frac{56 u}{3\ub} \ln^3(u)
- \frac{8 (1 - 31 \ub - 13 \ub^2 - 5 \ub^3)}{u^2\ub} \S_{1, 2}(\ub)
\no\\
&
+ \frac{8 (2 + \ub + 10 \ub^2 - \ub^3)}{u^2\ub} \ln(u) \Li_2(\ub)
- \frac{4 (3 - 4 \ub - 46 \ub^2 + 11 \ub^3)}{3u\ub^2} \ln^2(u)
\no\\
&
+ \bigg( \frac{4 (4 + 3 \ub - 72 \ub^2 - 7 \ub^3)}{3u^2\ub}
- \frac{12 (1 + \ub)^2 \pi^2}{u^3} \bigg) \Li_2(\ub)
- \frac{(359 + 362 \ub + 143 \ub^2)\pi^2}{18u^2}
\no\\
&
- \bigg( \frac{611 - 251 \ub}{9\ub}
- \frac{2 (3 + 32 \ub + 35 \ub^2 + 2 \ub^3)\pi^2}{3u^2\ub} \bigg) \ln(u)
- \frac43 \zeta_3 - \frac{3050}{27}
- 2 h_5(u),
\no\\
h_5(u) &=
h_2(u)
- \frac{4 (1 + \ub)}{3\ub} \Big( 12\calH_1(\ub) + \pi^2 \ln(2-u)\Big)
+ \frac{4 (12 - 21 \ub + 18 \ub^2 - 8 \ub^3)}{3u^3} \Li_3(\ub)
\no\\
&
+ \frac{2 (1 + 3 \ub^2)}{u^3}
\bigg( 8 \S_{2,2}(\ub) + 2 \Li_4(\ub)  - \Li_2(\ub)^2 + \frac{3 \pi^4}{20} \bigg)
+ \frac{16 (1 + \ub + \ub^2)}{3u^2} \ln(u)\Li_2(\ub)
\no\\
&
+ \frac{4 (14 + 15 \ub - 24 \ub^2 - 6 \ub^3)}{3u^3} \S_{1,2}(\ub)
- \bigg( \frac{472 u}{9\ub} - \frac{8 (1 - 2 \ub + 4 \ub^2) \pi^2}{3u^2\ub} \bigg) \ln(u)
\no\\
&
+ \bigg( \frac{4 (11 - 21 \ub + 4 \ub^2 - 5 \ub^3)}{3u^2\ub}
-  \frac{2 (1 + 3 \ub^2)\pi^2}{u^3} \bigg) \Li_2(\ub)
\no\\
&
- \frac{4 (2 - \ub) (5 - 8 \ub + 2 \ub^2)}{3u^3}
\bigg( \Li_3(-u) - \ln(u)\Li_2(-u) - \frac{\ln^2(u)+\pi^2}{2}\ln(1+u)\bigg)
\no\\
&
+ \frac{2 (13 - 18 \ub + 16 \ub^2)}{3u\ub} \ln^2(u)
- \frac{2 (13 + 4 \ub + 16 \ub^2)\pi^2}{9u^2}
- \frac{16}{u} \zeta_3
- \frac{5219}{54},
\no\\
h_6(u) &=
- \frac{8 u}{3\ub} \ln(u)
+ \frac{10}{3} \zeta_3
+ \frac{11\pi^2}{18}
- \frac{1381}{324},
\no\\
h_7(u) &=
h_3(u)
- \frac{16 (1 + \ub)^3}{3u^2\ub} \Li_2(\ub)
+ \frac{128 (1 + \ub + \ub^2)}{9u\ub} \ln(u)
+ \frac{32 (1 + \ub)\pi^2}{9u^2}
\no\\
&
- \frac{2 (251 + 325 \ub)}{27u},
\no\\
h_8(u) &=
- \frac{2 (3 + 20 \ub + 13 \ub^2)}{u^3}
\bigg( 8 \S_{2,2}(\ub) + 2 \Li_4(\ub)  - \Li_2(\ub)^2 + \frac{3 \pi^4}{20} \bigg)
- \frac{28 u}{3\ub} \ln^3(u)
\no\\
&
- \frac{2 (1 + 17 \ub + 51 \ub^2 + 3 \ub^3)}{u^2\ub} \Li_3(\ub)
+ \frac{4 (1 - 43 \ub - 93 \ub^2 - 9 \ub^3)}{u^2\ub} \S_{1, 2}(\ub)
\no\\
&
- \frac{8 (1 + 2 \ub + 15 \ub^2)}{u^2\ub} \ln(u) \Li_2(\ub)
+ \frac{9 + 13 \ub - 209 \ub^2 - 29 \ub^3}{3u\ub^2} \ln^2(u)
\no\\
&
+ \bigg( \frac{3 - 14 \ub - 84 \ub^2 + 402 \ub^3 + 125 \ub^4}{3u^2\ub^2}
+ \frac{2 (3 + 20 \ub + 13 \ub^2)\pi^2}{u^3} \bigg) \Li_2(\ub)
\no\\
&
- \bigg( \frac{81 - 539 \ub + 242 \ub^2}{18 \ub^2}
  + \frac{(1 + 17 \ub + 51 \ub^2 + 3 \ub^3)\pi^2}{u^2\ub} \bigg) \ln(u)
\no\\
&
+ \frac{2 (14 + 77 \ub + 17 \ub^2) \pi^2}{3u^2}
- \frac{11 - 3 \ub}{2\ub}
- 2 h_9(u),
\no\\
h_9(u) &=
\frac12 h_2(u) - \frac12 h_5(u)
- \frac{2 \ub (1 + 3 \ub)}{u^3}
\bigg( 8 \S_{2,2}(\ub) + 2 \Li_4(\ub)  - \Li_2(\ub)^2 + \frac{3 \pi^4}{20} \bigg)
\no\\
&
- \frac{2 (3 + 21 \ub - 24 \ub^2 + 2 \ub^3)}{3u^3} \Li_3(\ub)
- \frac{4 (1 + 7 \ub + 4 \ub^2)}{3u^2}\ln(u)\Li_2(\ub)
\no\\
&
- \frac{2 (1 + 69 \ub - 48 \ub^2 - 24 \ub^3)}{3u^3} \S_{1,2}(\ub)
+ \bigg( 4 + \frac{2 (1 - 11 \ub - 2 \ub^2)\pi^2}{3u^2} \bigg) \ln(u)
\no\\
&
- \bigg( \frac{2 (9 - 13 \ub - 18 \ub^2)}{3u^2}
- \frac{2 \ub (1 + 3 \ub) \pi^2}{u^3} \bigg) \Li_2(\ub)
- \frac{7 + 15 \ub}{3u} \ln^2(u)
\no\\
&
- \frac{2 (1 - 3 \ub + 6 \ub^2 - 2 \ub^3)}{3u^3}
\bigg( \Li_3(-u) - \ln(u)\Li_2(-u) - \frac{\ln^2(u)+\pi^2}{2}\ln(1+u)\bigg)
\no\\
&
+ \frac{(11 + 17 \ub + 38 \ub^2) \pi^2}{9u^2}
+ \frac{4 (3 - \ub)}{u} \zeta_3
- 4 \pi^2 \ln(2)
-\frac{5435}{108},
\no\\
h_{10}(u) &=
\frac12 h_3(u) - \frac12 h_7(u) - \frac{8 \pi^2}{9} + \frac{181}{27}.
\end{align}
Moreover, for the ratios $R_X$ in~(\ref{rfactors}) we need the following auxiliary functions,
\begin{align}
j_{1}(u) &= \frac {4 (u - 2) \left (u^2 + 2 u -
       2 \right) } {3 u^2 \ub} \, s_1(u)
       + \frac {16 \ub } {3 u^3} \, s_2(u)
       + \frac {8 (2 u - 7) \ub} {3 u^3} \, s_3(u) - \frac {4 u} {3 \ub} \, s_4(u)\no \\
       & + \frac {2 (u + 3) \left (u^2 - 1 \right)} {u^3} \, s_5(u)
       - \frac {16 (2 u - 3) \ub \, \Li_4 (\ub)} {u^3} - \frac {4
        \left (u^2 - 36 u + 25 \right) \Li_2 (\ub)} {u^2} \no \\
       &+ \frac {4 \left (4 u^3 - 8 u^2 + 5 u + 3 \right)
       (\Li_3 (\ub) - \zeta (3))} {u^3} + \frac {4 \left (7 u^3 +
       85 u^2 - 111 u + 3 \right) \Li_3 (u)} {u^3} \no \\
	&+ \frac {8 \pi ^2 (6 u - 17) \ub \, \Li_2 (\ub)} {3 u^3}+ \frac {6 u \, \Li_2 (\ub)} {\ub}
	- \frac {4 \left (5 u^3 + 63 u^2 - 83 u + 3 \right) \Li_2 (u) \ln(u)} {u^3} \no \\
        &  - \frac {2 \pi ^4 (20 u - 33) \ub} {45 u^3} - \frac {8 \left (u^2 + 51
           u - 62 \right) \zeta (3)} {u^2} - \frac {2 \pi ^2 \left (9 u^2 -
      73 u + 59 \right)} {3 u^2} \no \\
      & - \frac {32 \pi ^2 (5 u - 6) \ln(u)} {3 u^2} - \frac {2 \left (3 u^3 + 41 u^2 - 55 u +
       3 \right) \ln^2 (u) \ln(\ub)} {u^3} \no \\
      & + \frac {4 (2 u - 1)^2 \ln^2(u)} {\ub^2} + \frac {(9 u - 4) \ln(u)} {\ub} - \frac {2 (11 u - 25) \ln^2 (u)} {u} - 1 \, , \no \\
j_{2}(u) &=- \frac {2 (u - 2) \left (u^2 + 2 u - 2 \right)} {3 u^2 \ub} \, s_1(u)- \frac {8 \ub} {3 u^3} \, s_2 (u)
          + \frac {2 (4 u - 7)\ub} {3 u^3} \, s_3(u) \no \\
         &- \frac {(u + 3) \left (u^2 - 1 \right)} {u^3} \, s_5(u)
	 +\frac {2 \pi ^2 (12 u - 29) \ub \, \Li_2 (\ub)} {3 u^3} - \frac {4 (4 u - 15) \ub \, \Li_4 (\ub)} {u^3} \no \\
	 &- \frac {2 \left (25 u^2 -  99 u + 51 \right) \Li_2 (\ub)} {3 u^2}
	 - \frac {2 \left (3 u^3 + 29 u^2 - 37 u + 3 \right) (\Li_3 (\ub) - \zeta (3))} {u^3}  \no \\
	 &- \frac {2 \left (11 u^3 -   63 u^2 + 57  u + 3 \right) \Li_3 (u)} {u^3}
	 + \frac {2 \left (8 u^3 - 48 u^2 + 43 u + 3 \right) \Li_2 (u) \ln(u)} {u^3} \no \\
	 &- \frac {22 \Li_2 (\ub)} {3 \ub} - \frac {4 \pi ^4 (5 u - 18)\ub} {45 u^3} + \frac {2 \left (7 u^2 - 83 u + 86 \right) \zeta (3)} {u^2}
	 - \frac {\pi ^2 \left (8 u^2 - 69u + 67 \right)} {3 u^2} \no \\
	 &+ \frac {4 \pi ^2 \left (2 u^2 - 17 u + 18 \right) \ln(u)} {3 u^2} + \frac {\left (5 u^3 - 33 u^2 + 29 u + 3 \right) \ln^2 (u) \ln(\ub)} {u^3}
         - \frac {13 \ln^2 (u)} {3 \ub} \no \\
         &- \frac {4 \pi ^2 \ln(u)} {3\ub} + \frac {203 \ln(u)} {9 \ub} - \frac {(8 u - 51) \ln^2 (u)} {3 u}
	 - \frac {257 \ln(u)} {9} + \frac{269}{9} \, , \no \\
j_{3}(u) &= -\frac{26}{9} \, g_9(u) + \frac{8u}{3\ub} \, \left[\ln^2 (u)+\Li_2 (\ub)\right] - \frac{76}{9} \, , \no \\
j_{4}(u) &=\frac{32 \pi ^2 (u+2) \ub}{9 u^2}+\frac{32 \ub (\Li_3(\ub)-\zeta (3))}{u^3}-\frac{8 (u-2) \left(u^2+2 u-2\right)
   \Li_2(\ub)}{3 u^2 \ub} \no \\
   &-\frac{52 u \ln(u)}{9\ub}-\frac{104}{3 u}+\frac{80 \ln(u)}{3 u}+\frac{236}{9} \, , \no \\
j_{5}(u) &= -\frac {4 (u - 2) \left (u^2 - 2 u +
       2 \right) } {3 u^2 \ub} \, s_1(u)
       + \frac {16 \ub } {3 u^3} \, s_2(u)
       - \frac {8 (2u^2-12u+11)} {3 u^3} \, s_3(u) \no \\
       & + \frac {4} {3 \ub} \, s_4(u)- \frac {2 (u + 1) \left (u^2 +2u+7 \right)} {3u^3} \, s_5(u)
       + \frac {16 (2 u^2-8u+7) \Li_4 (\ub)} {u^3}  \no \\
       &- \frac {2 \left (51u^2 - 328 u + 250 \right) \Li_2 (\ub)} {3u^2}
       - \frac {4 \left (u^3 + 48 u^2 - 69 u - 7 \right)
       (\Li_3 (\ub) - \zeta (3))} {3u^3}  \no \\
	&+ \frac {4 \left (3 u^3 + 315 u^2 - 519 u + 7 \right) \Li_3 (u)} {3u^3}
       - \frac {8 \pi ^2 (6 u^2 -32 u +29) \Li_2 (\ub)} {3 u^3}- \frac {2 \Li_2 (\ub)} {\ub} \no \\
        &  + \frac {4 \left (2 u^3 - 237 u^2 + 387 u - 7 \right) \Li_2 (u) \ln(u)} {3u^3}
	+ \frac {2 \pi ^4 (20 u^2 - 83u+73)} {45 u^3}  \no \\
      & + \frac {8 (4 u^2 - 183 u + 306) \zeta (3)} {3u^2}
	   - \frac {\pi^2 (11 u^2 - 206 u + 218 )} {3 u^2}
      + \frac {8 \pi^2 (u^2 - 72 u + 120) \ln(u)} {9 u^2}  \no \\
      & + \frac {2 \left (7 u^3 - 159 u^2 + 255 u -
       7 \right) \ln^2 (u) \ln(\ub)} {3u^3}+ \frac {4 \ln^2(u)} {\ub} + \frac {\ln(u)} {\ub} -13\ln(u) \no \\
      & - \frac {2 (27 u - 125) \ln^2 (u)} {3u} -8\pi^2\ln(2)+\frac{563}{24}\, , \no \\
j_{6}(u) &= \frac {2 (u - 2) \left (u^2 - 2 u +
       2 \right) } {3 u^2 \ub} \, s_1(u)
       - \frac {8 \ub } {3 u^3} \, s_2(u)
       - \frac {2 (7u^2-17u+11)} {3 u^3} \, s_3(u) \no \\
       & + \frac {(u + 1) \left (u^2 +2u+7 \right)} {3u^3} \, s_5(u)
       + \frac {4 (7 u^2-25u+19) \Li_4 (\ub)} {u^3}  \no \\
       &- \frac {2 \left (35u^2 - 133 u + 73 \right) \Li_2 (\ub)} {3u^2}
       + \frac {2 \left (7u^3 - 99 u^2 + 129 u - 7 \right)
       (\Li_3 (\ub) - \zeta (3))} {3u^3}  \no \\
	&- \frac {2 \left (45 u^3 - 273 u^2 + 273 u + 7 \right) \Li_3 (u)} {3u^3}
       - \frac {2 \pi ^2 (21 u^2 -59 u +41) \Li_2 (\ub)} {3 u^3}+ \frac {22 \Li_2 (\ub)} {3\ub} \no \\
        &  + \frac {2 \left (28 u^3 - 204 u^2 + 207 u + 7 \right) \Li_2 (u) \ln(u)} {3u^3}
	+ \frac {\pi ^4 (35 u^2 - 122 u+92)} {45 u^3}  \no \\
      & + \frac {2 (40 u^2 \!- 369 u + 378) \zeta (3)} {3u^2}
	   - \frac {\pi^2 (68 u^2 - 279 u + 267 )} {9 u^2}
      + \! \frac {4 \pi^2 (8u^2 - 75 u + 78) \! \ln(u)} {9 u^2}  \no \\
      & + \frac {\left (11 u^3 - 135 u^2 + 141 u + 7 \right) \ln^2 (u) \ln(\ub)} {3u^3}
      + \frac {13 \ln^2(u)} {3\ub} - \frac {269\ln(u)} {9\ub} + \frac {4\pi^2 \ln(u)} {3\ub} \no \\
      & +\frac{215\ln(u)}{9}- \frac { (40 u - 73) \ln^2 (u)} {3u} +4\pi^2\ln(2)-\frac{4421}{216}\, , \no \\
j_{7}(u) &=\frac{38}{9} \, g_9(u) - \frac{8u}{3\ub} \, \left[\ln^2 (u)+\Li_2 (\ub)\right] + \frac{4\pi^2}{9}+ \frac{205}{54}\, , \no \\
j_{8}(u) &=-\frac{8 \pi ^2 (u^2+8u-16)}{9 u^2}+\frac{32 \ub (\Li_3(\ub)-\zeta (3))}{u^3}+\frac{8 (u-2) \left(u^2-10 u+10\right)
   \Li_2(\ub)}{3 u^2 \ub} \no \\
   &+\frac{76 \ln(u)}{9\ub}-\frac{232}{3 u}-\frac{4 (19u-156) \ln(u)}{9 u}+\frac{1429}{54} \, , \no \\
j_{9}(u) &= -\frac{5 \pi ^2 (5u+4)}{6 u}-\frac{16 \ub (\Li_3(\ub)-\zeta (3))}{u^2}-\frac{2 (u-2)
   \Li_2(\ub)}{u} +\frac{u^2 \ln^2(u)}{\ub^2}\no \\
   &- 12 \ln(u)-6\zeta(3)+4\pi^2\ln(2)+\frac{563}{16} \, , \no \\
j_{10}(u) &= \frac{4 \pi ^2 (u+1)}{3 u}+\frac{8 \ub (\Li_3(\ub)-\zeta (3))}{u^2}+ 4 \ln(u)+3\zeta(3)-2\pi^2\ln(2)-\frac{5141}{144} \, ,
\end{align}
with
\begin{align}
s_1(u) &=12 \calH_1(\ub) + \pi ^2 \ln(2 - u) \, ,  \no \\
s_2(u) &=12 \calH_2(\ub) - \pi ^2 \Li_2 (-\ub)  \, , \no \\
s_3(u) &=3 \Li_2^2 (\ub)- 24 \S_{2,2}(\ub) - \frac {17 \pi ^4} {60} \, , \no \\
s_4(u) &=6 \Li_3 (u) - 3 \Li_2 (u) \ln(u) + 3 \Li_3 (\ub) - 2 \pi ^2 \ln(u) - 6 \zeta (3)  \, , \no \\
s_5(u) &=-2 \Li_3 (-u) + 2 \Li_2 (-u) \ln(u) + \ln(u + 1) \ln^2 (u) + \pi ^2 \ln(u + 1) \, .
\end{align}
\end{appendix}


\end{document}